\documentclass[a4paper,11pt]{article}
\pdfoutput=1 

\usepackage{jcappub} 

\usepackage[T1]{fontenc} 

\title{Probing pre-inflationary anisotropy with directional variations in the gravitational wave background}

\author{Yu Furuya,}
\author{Yuki Niiyama}
\author{and Yuuiti Sendouda}
\affiliation{
Graduate School of Science and Technology, Hirosaki University,\\
3 Bunkyocho, Hirosaki, Aomori 036-8561, Japan
}

\emailAdd{furuya@tap.st.hirosaki-u.ac.jp}
\emailAdd{niiyama@tap.st.hirosaki-u.ac.jp}
\emailAdd{sendouda@hirosaki-u.ac.jp}

\abstract{We perform a detailed analysis on a primordial gravitational-wave background amplified during a Kasner-like pre-inflationary phase allowing for general triaxial anisotropies.
It is found that the predicted angular distribution map of gravitational-wave intensity on large scales exhibits topologically distinctive patterns according to the degree of the pre-inflationary anisotropy, thereby serving as a potential probe for the pre-inflationary early universe with future all-sky observations of gravitational waves.
We also derive an observational limit on the amplitude of such anisotropic gravitational waves from the $ B $-mode polarisation of the cosmic microwave background.}


\begin{document}
\maketitle
\flushbottom

\section{Introduction}

There has been an accumulation of observational evidence for cosmic inflation \cite{Guth:1980zm,Sato:1980yn} in the past decades, one of the most crucial being the precise measurements of the cosmic microwave background (CMB) \cite{Ade:2015lrj}.
Among others, a subject currently attracting a particular attention in this field is the search for the primordial gravitational waves (PGWs) emerging from the quantum nature of space-time during an early accelerated expansion \cite{Mukhanov:1981xt}, which is thought to be a \emph{direct} evidence of the occurrence of inflation.
Direct detections of such PGWs in low-frequency bands are indeed one of the primary aims of future planned laser-interferometer experiments in space, such as eLISA \cite{AmaroSeoane:2012je}, DECIGO \cite{Seto:2001qf}, and Big-Bang Observer \cite{Crowder:2005nr}.

There are, however, still many challenges to overcome in laser-interferometer experiments in spite of the recent success of the first observation of gravitational waves from coalescing binary black holes \cite{Abbott:2016blz}.
As an alternative approach, there have been attempts at measuring the $ B $-mode polarisations of the CMB as an imprint of the PGWs \cite{Seljak:1996gy,Kamionkowski:1996zd,Kamionkowski:1996ks}.
Past and ongoing projects for CMB $ B $-mode polarisation measurement at low multipoles include POLARBEAR \cite{Ade:2014afa}, ACTpol \cite{Naess:2014wtr}, BICEP2/Keck array and Planck \cite{Ade:2015tva} (see also \cite{Ade:2014xna}), and SPTpol \cite{Keisler:2015hfa}.
Many future experiments are also planned, such as POLARBEAR-2 and Simons Array \cite{Suzuki:2015zzg}, and LiteBIRD \cite{Matsumura:2013aja}.
Several constraints on tensor perturbations from the current bound on the CMB $ B $-mode have been obtained in \cite{Ade:2015tva,Ade:2015xua}.

One of the goals beyond confirming the occurrence of cosmic inflation would be determination of the initial condition of inflation.
If there are earlier stages preceding the onset of inflation, it is well anticipated that there were anisotropies and/or inhomogeneities of order the energy scale of unified theories such as Grand Unification Theories (GUTs) or superstrings.
Investigations of early anisotropies may bring us useful information to construct the theory of elementary particles beyond the Standard Model and even quantum gravity.
This serves as another motivation for investigating the anisotropic stages before an isotropic inflation.

Then, what is to be understood is how the universe has evolved into the currently observed homogeneous and isotropic state.
A key clue to this issue would be Wald's cosmic no-hair theorem \cite{Wald:1983ky}, stating that the Bianchi models for homogeneous anisotropic universe \cite{Ellis:1968vb} inevitably evolve towards isotropic de Sitter space in the presence of a large enough positive cosmological constant $ \Lambda $\,.

It is generically expected that anisotropies of cosmological expansion would have great impacts on the evolution of cosmological perturbations.
Indeed, several observational signatures of an anisotropic pre-inflation were discussed in \cite{Gumrukcuoglu:2007bx,Gumrukcuoglu:2008gi,Pitrou:2008gk}.
Among others, a remarkable finding is that amplification of gravitational waves occurs during the pre-inflationary Kasner regime \cite{Gumrukcuoglu:2008gi,Kofman:2011tr}, whose efficiency varies with the direction in the sky.
In particular, in \cite{Gumrukcuoglu:2008gi}, G{\"{u}}mr{\"{u}}k\c{c}{\"{u}}o\u{g}lu et al.\ investigated such gravitational waves from an inflationary background driven by a scalar field.

The purpose of the present paper, therefore, is to give further insights into the connection between direction-dependent gravitational waves and primordial pre-inflationary anisotropies.
While G{\"{u}}mr{\"{u}}k\c{c}{\"{u}}o\u{g}lu et al.'s analysis in \cite{Gumrukcuoglu:2008gi} was restricted to a particular background with axisymmetry, in contrast, we treat general \emph{triaxial} backgrounds.
To do so, we take the so-called Kasner-de Sitter metric as a simpler background in which initially anisotropic expansion of space is isotropised due to a cosmological constant rather than a scalar field.
As analysed in \cite{Gumrukcuoglu:2008gi,Kofman:2011tr}, the amplification of gravitational waves originates from an instability in the tensor sector and we expect even our simple model captures its essential features.
This simple model also allows us to generate the all-sky map of gravitational-wave intensity from which we could decode the degree of the primordial anisotropy.
Also we give some tentative constraints on the model parameters from the current bounds on the CMB $ B $-mode polarisation.

The organisation of the paper is as follows.
In section~\ref{sec:basic}, we describe the basic equations for the background and perturbations.
In particular, we employ a gauge-invariant formulation for cosmological perturbations in the Bianchi type-I universe recently constructed by Pereira et al.\ \cite{Pereira:2007yy} (see also \cite{Tomita:1985me} for an earlier attempt).
In section~\ref{sec:gws}, we show how gravitational waves evolve depending on the direction of wavevectors as well as on the anisotropy of the background universe.
In section~\ref{sec:limit}, we demonstrate how limits can be given to the initial anisotropy and the initial power spectrum of the pre-inflationary gravitational waves from future all-sky observations of the PGWs.
Finally, in section~\ref{sec:concl}, we conclude.

Throughout the paper we use natural units with $ c = \hbar = k_\mathrm B = 1 $.
The Latin indices $ i,j $ are spatial and run through $ 1,2,3 $.
We will use the arrow notation to denote spatial contravariant vectors such as $ \vec V \equiv (V^1,V^2,V^3) $.

\section{\label{sec:basic}Basic equations}

\subsection{Kasner-de Sitter inflation}

Hereafter we consider the so-called Kasner-de Sitter (KdS) metric as a simple model for an inflationary universe with an initial anisotropy.
It is an exact solution to the Einstein equations in the presence of a positive cosmological constant $ \Lambda $\,, whose line element is given by
\begin{equation}
g_{\mu\nu}\,\mathrm dx^\mu\,\mathrm dx^\nu
= -\mathrm dt^2
  + a_\mathrm{iso}^2\,\sinh^{2/3}\left(3Ht\right)\,
    \sum_{i=1}^3 \tanh^{2q_i}\left(\frac{3Ht}{2}\right)\,(\mathrm dx^i)^2\,,
\label{eq:metric}
\end{equation}
where $ a_\mathrm{iso} $ is an arbitrary positive constant, $ H \equiv \sqrt{\Lambda/3} $\,, and the three exponents $ q_i $ ($ i = 1,2,3 $) satisfy the constraints
\begin{equation}
\sum_{i=1}^3 q_i
= 0\,,
\quad
\sum_{i=1}^3 q_i^2
= \frac{2}{3}\,.
\end{equation}
The KdS metric is classified into the Bianchi type-I cosmological model \cite{Ellis:1968vb}.
We define the average scale factor $ a(t) $ and the spatial metric $ \gamma_{ij}(t) $ by
\begin{equation}
a(t)
\equiv
  a_\mathrm{iso}\,\sinh^{1/3}\left(3Ht\right)\,,
\quad
\gamma_{ij}(t)
\equiv
  \tanh^{2q_i}\left(\frac{3Ht}{2}\right)\,\delta_{ij}\,.
\end{equation}
At earlier times with $ t \ll H^{-1} $\,, the metric behaves like a vacuum Kasner solution
\begin{equation}
a(t)
\simeq
  a_\mathrm{iso}\,(3Ht)^{1/3}\,,
\quad
\gamma_{ij}(t)
\simeq
  \left(\frac{3Ht}{2}\right)^{2q_i}\,\delta_{ij}\,,
\end{equation}
while at later times with $ t \gg H^{-1} $\,, the metric asymptotes to a de Sitter solution,
\begin{equation}
a(t)
\simeq
  2^{-1/3}\,a_\mathrm{iso}\,\mathrm e^{Ht}\,,
\quad
\gamma_{ij}(t)
\simeq
  \delta_{ij}\,.
\end{equation}
Hence the KdS metric describes a universe evolving from an anisotropic initial phase to an isotropic inflation driven by $ \Lambda $\,.
The isotropisation is achieved around the time $ t_\mathrm{iso} \equiv H^{-1} $\, and the constant $ a_\mathrm{iso} $ serves as the normalisation of the scale factor at the beginning of the isotropic de Sitter inflation.
The duration of isotropic de Sitter expansion in this model is necessarily bounded from above.

This simple (pre-)inflationary model has some flaws as a realistic setup for the early universe such as lacking of mechanisms to realise a decline of the expansion rate and to exit from the inflationary phase, i.e., reheating.
Precisely, what we assume is that the superhorizon fluctuations reentering the cosmological horizons at late times after reheating have exited the horizon at the early era when the metric is well approximated by \eqref{eq:metric}.
Since the presence of global anisotropies on cosmological scales is observationally not favoured, we should require $ a_\mathrm{iso}\,H \lesssim a(t_0)\,H(t_0) $\,, where $ t_0 $ is the present age of the universe.
Note that this condition is equivalent to the one for evading the horizon and flatness problems.

\begin{figure}[htbp]
\begin{center}
\includegraphics[scale=0.7]{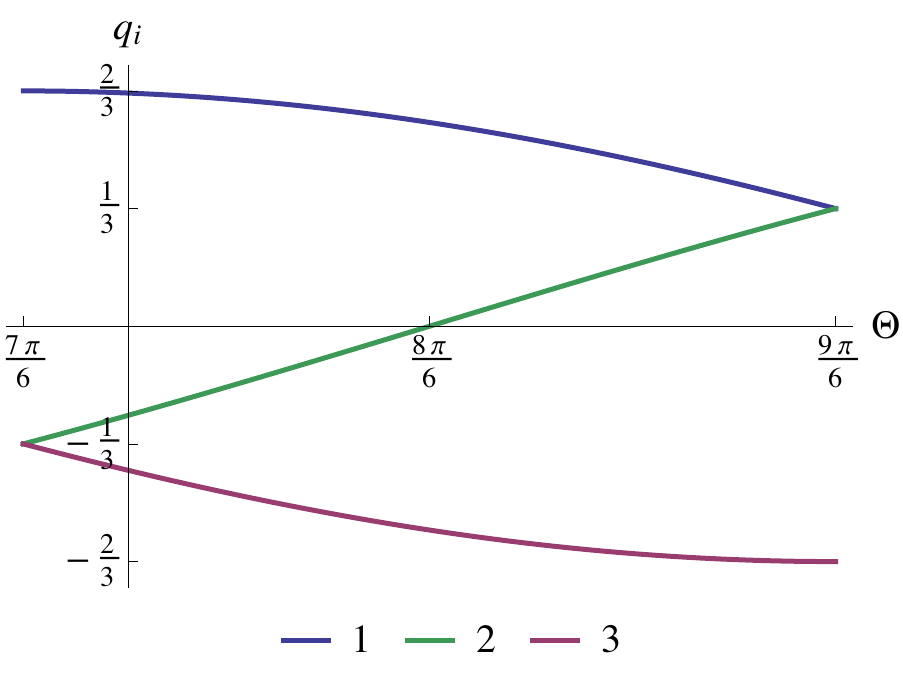}
\end{center}
\caption{\label{fig:q}Values of $ q_i $ for $ \frac{7\pi}{6} \leq \Theta \leq \frac{9\pi}{6} $\,.}
\end{figure}

It is useful to introduce an angular parameter $ \Theta $ to express the exponents $ q_i $ as
\begin{equation}
q_1
= \frac{2}{3}\,\sin\left(\Theta-\frac{2\pi}{3}\right)\,,
\quad
q_2
= \frac{2}{3}\,\sin\left(\Theta-\frac{4\pi}{3}\right)\,,
\quad
q_3
= \frac{2}{3}\,\sin\Theta\,.
\end{equation}
The parameter $ \Theta $ quantifies the degree of anisotropy of pre-inflationary cosmological expansion.
With no loss of generality, we can restrict our attention to a range $ \frac{7\pi}{6} \leq \Theta \leq \frac{9\pi}{6} $\,, within which the order of the exponents is $ q_1 \geq q_2 \geq q_3 $\,, see figure~\ref{fig:q}.
Any KdS metric is equivalent, up to relabelling the axes and/or reversing their orientations, to one with the value of $ \Theta $ lying in the above range.
The expansion of the universe is axisymmetric for $ \Theta = \frac{7\pi}{6} $ ($ (q_1,q_2,q_3) = (\frac{2}{3},-\frac{1}{3},-\frac{1}{3}) $) and for $ \Theta = \frac{9\pi}{6} $ ($ (q_1,q_2,q_3) = (\frac{1}{3},\frac{1}{3},-\frac{2}{3}) $).
In the following analysis we shall often take $ \Theta = \frac{8\pi}{6} $ as a fiducial value, for which the anisotropy exponents are $ (q_1,q_2,q_3) = (\frac{1}{\sqrt 3},0,-\frac{1}{\sqrt 3}) $\,.

The KdS metric has an initial Kasner singularity at $ t = 0 $ unless $ \Theta = \frac{7\pi}{6} $\,, so we necessarily introduce an initial time $ t_\mathrm{ini} > 0 $\,, which is supposed to be before $ t_\mathrm{iso} $\,.
We do not discuss the initial singularity problem, but for a possible resolution, see \cite{Blanco-Pillado:2015dfa}.
If inflation occurred at the GUT scale with $ H \sim H_\mathrm{GUT} \sim 10^{-6}\,m_\mathrm{Pl} $\,, then $ t_\mathrm{ini} > t_\mathrm{Pl} \sim 10^{-6}\,H_\mathrm{GUT}^{-1} $ is required so as to avoid the trans-Planckian problems.
In what follows, we take $ t_\mathrm{ini} = 10^{-5}\,t_\mathrm{iso} = 10^{-5}\,H^{-1} $ as a fiducial value for the initial time.

The conformal time $ \eta $ is defined by $ \mathrm d\eta = \mathrm dt/a(t) $\,.
A prime will denote differentiation with respect to $ \eta $\,.
The cosmic expansion is characterised by the average Hubble parameter $ \mathcal H $ and the shear tensor $ \sigma_{ij} $ defined respectively as
\begin{equation}
\mathcal H
\equiv
  \frac{a'}{a}
= a_\mathrm{iso}\,H\,\frac{\cosh(3Ht)}{\sinh^{2/3}(3Ht)}\,,
\quad
\sigma_{ij}
\equiv
  \frac{1}{2}\,\gamma_{ij}'
= 3 q_i\,a_\mathrm{iso}\,H\,
  \frac{\tanh^{2q_i}\left(\frac{3Ht}{2}\right)}
       {\sinh^{2/3}(3Ht)}\,
  \delta_{ij}\,.
\end{equation}

The inverse of $ \gamma_{ij} $ is defined by $ \gamma^{ik}\,\gamma_{kj} = \delta^i{}_j $\,.
The indices of spatial tensors will be raised and lowered by $ \gamma^{ij} $ and $ \gamma_{ij} $ such as $ \sigma^i{}_j \equiv \gamma^{ik}\,\sigma_{kj} $\,.

\subsection{Polarisation basis}

As usual, we will parameterise the harmonic modes of waves by a set of constants $ (k_1,k_2,k_3) $, which are regarded as the covariant components of a wavevector in the $ (x^1,x^2,x^3) $ comoving coordinate frame.
What is unusual is that, on an anisotropic KdS background, the contravariant components $ k^i \equiv k_j\,\gamma^{ij} $ are not constant.
The contravariant wavevector $ \vec k \equiv (k^1,k^2,k^3) $ changes its direction and norm, $ \sqrt{k_i\,k^i} $\,, during the anisotropic Kasner regime, $ t \lesssim t_\mathrm{iso} $\,, and after the universe is isotropised, $ t \gtrsim t_\mathrm{iso} $\,, it comes to coincide with its covariant dual as $ \lim_{t\to\infty} \vec k = (k_1,k_2,k_3) $\,.

It is useful to introduce time-dependent polar angles $ (\beta,\gamma) $ to parameterise the normalised wavevector $ \vec{\hat k} \equiv \vec k/\sqrt{k_i\,k^i} $ as \cite{Pitrou:2008gk}
\begin{equation}
\vec{\hat k}
\equiv
\begin{pmatrix}
\mathrm e^{-\beta_1}\,\sin\beta\,\cos\gamma \\
\mathrm e^{-\beta_2}\,\sin\beta\,\sin\gamma \\
\mathrm e^{-\beta_3}\,\cos\beta
\end{pmatrix}\,,
\label{eq:k}
\end{equation}
where we have introduced the notation
\begin{equation}
\mathrm e^{-\beta_i}
\equiv
  \sqrt{\gamma^{ii}}
= \tanh^{-q_i}\left(\frac{3Ht}{2}\right)\,.
\label{eq:beta}
\end{equation}
The orthonormal polarisation vector basis perpendicular to $ \vec{\hat k} $ is introduced as
\begin{equation}
\begin{aligned}
\vec e_{(1)}
&
\equiv
\begin{pmatrix}
e^{-\beta_1}\,
(\cos\beta\,\cos\gamma\,\cos\alpha - \sin\gamma\,\sin\alpha) \\
e^{-\beta_2}\,
(\cos\beta\,\sin\gamma\,\cos\alpha + \cos\gamma\,\sin\alpha) \\
-e^{-\beta_3}\,\sin\beta\,\cos\alpha
\end{pmatrix}\,, \\
\vec e_{(2)}
&
\equiv
\begin{pmatrix}
-e^{-\beta_1}\,
(\cos\beta\,\cos\gamma\,\sin\alpha + \sin\gamma\,\cos\alpha) \\
-e^{-\beta_2}\,
(\cos\beta\,\sin\gamma\,\sin\alpha - \cos\gamma\,\cos\alpha) \\
e^{-\beta_3}\,
\sin\beta\,\sin\alpha
\end{pmatrix}\,,
\end{aligned}
\end{equation}
where the arbitrary angle $ \alpha $ represents the rotation degree of freedom of $ (\vec e_{(1)},\vec e_{(2)}) $ around $ \vec{\hat k} $\,.
As in \cite{Pitrou:2008gk}, we determine $ \alpha $ by imposing a condition
\begin{equation}
\alpha'
= -\gamma'\,\cos\beta\,.
\label{eq:Euler}
\end{equation}
Accordingly, the orthonormal tensor basis is defined in terms of the vector basis as
\begin{equation}
\varepsilon^+_{ij} 
\equiv
  \frac{e^{(1)}_i\,e^{(1)}_j - e^{(2)}_i\,e^{(2)}_j}{\sqrt 2}\,,
\quad
\varepsilon^\times_{ij}
\equiv
  \frac{e^{(1)}_i\,e^{(2)}_j + e^{(2)}_i\,e^{(1)}_j}{\sqrt 2}\,.
\end{equation}
Using the polarisation basis, the shear tensor is decomposed into the scalar, vector, and tensor components as
\begin{equation}
\sigma^{(\mathrm S)}
\equiv
  \sigma_{ij}\,\hat k^i\,\hat k^j\,,
\quad
\sigma^{(\mathrm V)}_{(a)}
\equiv
  \sigma_{ij}\,\hat k^i\,e_{(a)}^j\,,
\quad
\sigma^{(\mathrm T)}_\lambda 
\equiv
  \sigma_{ij}\,\varepsilon_\lambda^{ij}\,,
\end{equation}
where $ a = 1,2 $ for the vector and $ \lambda = +,\times $ for the tensor components, respectively.

\subsection{Gravitational wave equations}

In order to analyse evolution of gravitational waves on a KdS spacetime, we apply the gauge-invariant formulation of cosmological perturbations in Bianchi type-I models developed by Pereira et al.\ \cite{Pereira:2007yy}.
In this formalism, the general perturbed metric is given by
\begin{equation}
(g_{\mu\nu} + \delta g_{\mu\nu})\,\mathrm dx^{\mu}\,\mathrm dx^{\nu}
= a(\eta)^2\,
  \left[
   -(1 + 2A)\,\mathrm d\eta^2
   + 2 B_i\,\mathrm dx^i\,\mathrm d\eta
   + (\gamma_{ij} + h_{ij})\,\mathrm dx^i\,\mathrm dx^j
  \right]\,,
\end{equation}
where the $ (0i) $ and $ (ij) $ components are respectively decomposed into the scalar, vector, and tensor variables as
\begin{equation}
B_i
= \partial_i B + \bar B_i\,,
\quad 
h_{ij}
= 2\,(\gamma_{ij} + \mathcal H^{-1}\,\sigma_{ij})\,C
  + 2 \partial_{ij}E + 2 \partial_{(i}E_{j)} + 2E_{ij}\,,
\end{equation}
where the vector and tensor variables satisfy $ \partial_i \bar B^i = \partial_i E^i = \partial_i E^i{}_j = E^i{}_i = 0 $\,.
In vacuum (plus a cosmological constant), the only dynamical degrees of freedom  are represented by the two polarisation components of the gauge-invariant tensor variable $ E_{ij} $ \cite{Pereira:2007yy}, defined using the tensor polarisation basis as
\begin{equation}
E_\lambda
\equiv
  \int\!\frac{\mathrm d^3x}{(2\pi)^{3/2}}\,\mathrm e^{-\mathrm i k_l x^l}\,
  \varepsilon_\lambda^{ij}\,E_{ij}
\quad
(\lambda=+,\times)\,.
\end{equation}
Introducing $ \mu_\lambda \equiv a\,E_\lambda $ for convenience, the equations of motion for gravitational waves are given by \cite{Pereira:2007yy}
\begin{equation}
\begin{aligned}
&
\mu_+'' + \omega_+^2\,\mu_+ + \xi\,\mu_\times = 0\,, \\
&
\mu_\times'' +\omega_\times^2\,\mu_\times + \xi\,\mu_+ = 0\,,
\end{aligned}
\label{eq:eom}
\end{equation}
where
\begin{equation}
\begin{aligned}
\omega_+^2
&
\equiv
  k_i\,k^i
  - \frac{a''}{a}
  - \frac{\left(a^2\,\sigma^{(\mathrm S)}\right)'}{a^2}
  - 2\,\left(\sigma_\times^{(\mathrm T)}\right){}^2
  - \frac{2}{a^2}\,\left(
     \frac{a^2\,\left(\sigma_+^{(\mathrm T)}\right){}^2}
          {2 \mathcal H-\sigma^{(\mathrm S)}} 
    \right)'\,, \\
\omega_\times^2
&
\equiv
  k_i\,k^i
  - \frac{a''}{a}
  - \frac{\left(a^2\,\sigma^{(\mathrm S)}\right)'}{a^2}
  - 2\,\left(\sigma_+^{(\mathrm T)}\right){}^2
  - \frac{2}{a^2}\,\left(
     \frac{a^2\,\left(\sigma_\times^{(\mathrm T)}\right){}^2}
          {2 \mathcal H-\sigma^{(\mathrm S)}} 
    \right)'\,,
\end{aligned}
\label{eq:omega}
\end{equation}
and
\begin{equation}
\xi
\equiv 
  2 \sigma_+^{(\mathrm T)}\,\sigma_\times^{(\mathrm T)}
  - \frac{2}{a^2}\,\left( 
     \frac{a^2\,\sigma_+^{(\mathrm T)}\,\sigma_\times^{(\mathrm T)}}
          {2 \mathcal H-\sigma^{(\mathrm S)}} 
    \right)'\,.
\label{eq:xi}
\end{equation}

One could guess from the expressions for ``squared frequencies'', $ \omega_\lambda^2 $\,, that the shear can give negative contributions and might lead to some instabilities.
Indeed, it will be revealed that the contributions from the shear control how gravitational waves grow during the pre-inflationary anisotropic expansion.
In the sequel, we will pay particular attention to how much and how long $ \omega_\lambda^2 $ receives negative contributions from the shear, which crucially depends on the values of $ (k_1,k_2,k_3) $.

The $ \xi $ term characterises the interaction between the two polarisation modes.
Occurrence of interactions in general makes the analyses more complicated, but eq.~\eqref{eq:xi} implies that $ \xi $ vanishes if either of $ \sigma^{(\mathrm T)}_\lambda $ is zero.
We will take full advantage of this property in the following analysis.

\section{\label{sec:gws}Angular anisotropies of gravitational wave background}

In this section we analyse time evolutions of gravitational waves labelling them with the \emph{final} wavevector $ \lim_{t\to\infty} \vec k = (k_1,k_2,k_3) $\,.
The main object of interest in our current study is the distribution of gravitational-wave intensity on a sphere of radius $ k_0 \equiv \sqrt{k_1^2+k_2^2+k_3^2} $ defined in the $ k $-space, which, after being reflected through the origin and mapped to the celestial sphere, will provide an all-sky map of a gravitational wave background which we will be able to observe at a comoving wavenumber $ k_0 $\,.

Since there is a natural correspondence between the $ k $-space and the position space, we sometimes mix up them with each other and even use common terminologies.
For instance, a term ``principal axes of expansion'' may refer to both the $ x^i $-axes in the position space and the $ k^i $-axes in the $ k $-space depending on the context.

Since we are interested in gravitational waves currently on large scales and since we require $ a_\mathrm{iso}\,H \lesssim a(t_0)\,H(t_0) $\,, we only consider modes which had a shorter wavelength than the Hubble radius at the isotropisation time $ t_\mathrm{iso} = H^{-1} $\,.
We will take $ k_0 = \mathcal O(10\textnormal{--}10^3) \times a_\mathrm{iso}\,H $ as reference values for the wavenumber.

\subsection{\label{sec:k_dir}Rotation of wavevector during anisotropic expansion}

\begin{figure}[htbp]
\begin{center}
\includegraphics[scale=0.6]{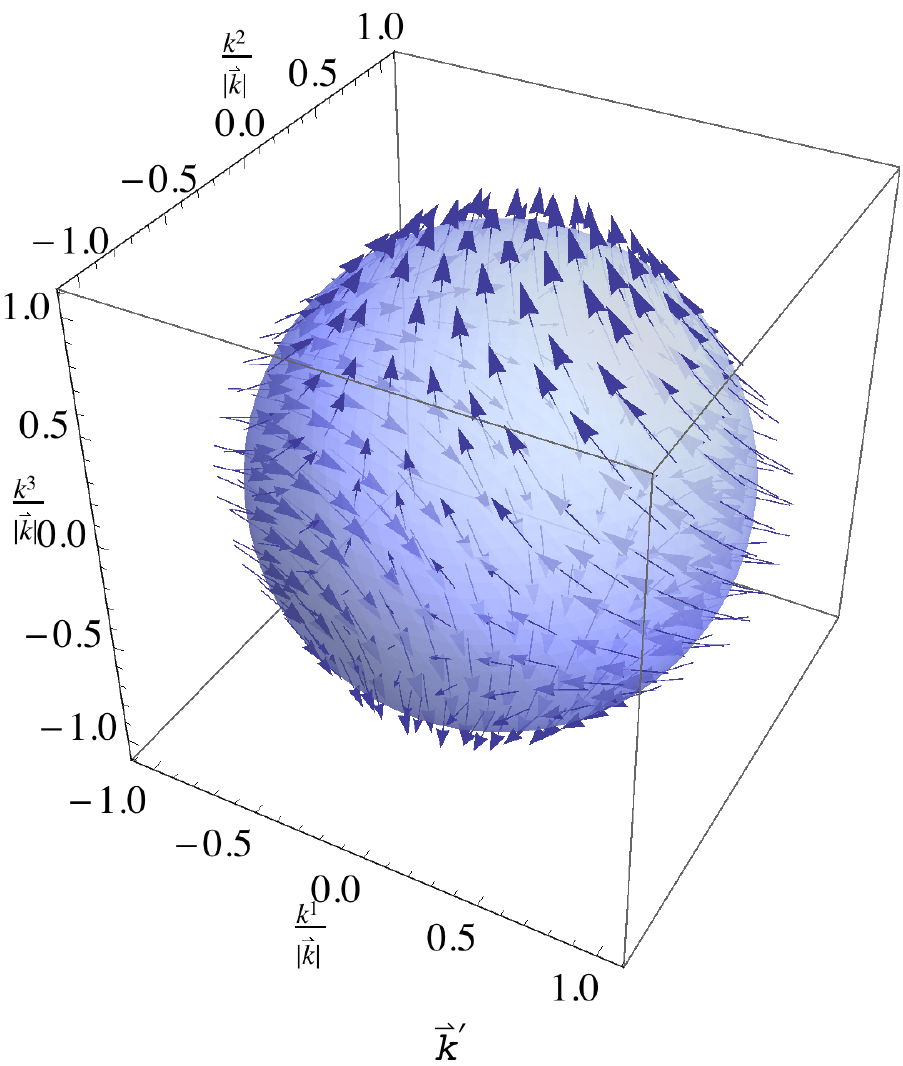}
\includegraphics[scale=0.6]{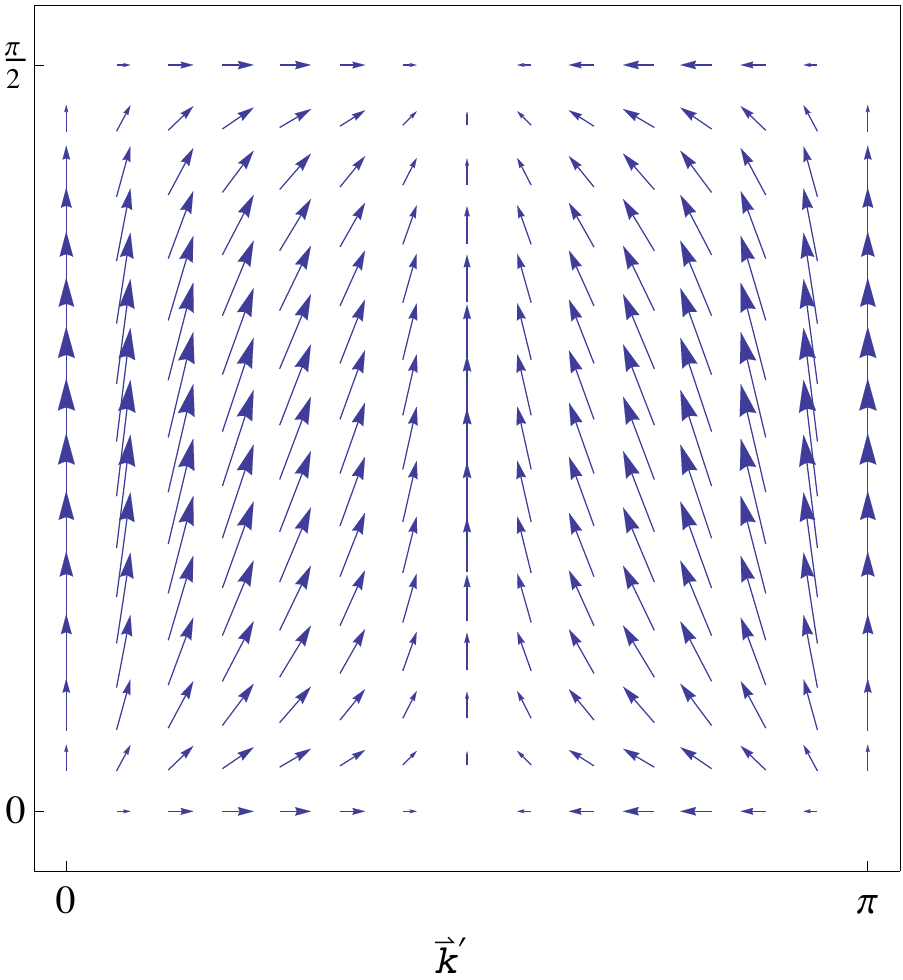}
\end{center}
\caption{\label{fig:k_dir}Left: A phase portrait of the time derivative $ \vec k' $ in the $ k $-space.
Right: The Mercator projection of the half hemisphere specified by $ k^1 \geq 0 $ and $ k^3 \geq 0 $.
The background is characterised by the anisotropy parameter $ \Theta = \frac{8\pi}{6} $\,.}
\end{figure}

First we explain how a wavevector evolves with time.
Figure~\ref{fig:k_dir} is a portrait of the time derivatives of $ \vec k $ drawn on the unit sphere in the $ k $-space (left) and the Mercator projection of a portion of the sphere (right).
As implied by eqs.~\eqref{eq:k} and \eqref{eq:beta}, the general trend is that a wavevector $ \vec k $ changes its direction towards the principal axis of the slowest expansion of the three, i.e., with the smallest exponent $ q_i $\,.
Since $ q_3 $ is always the smallest in our setup (except for $ \Theta = \frac{7\pi}{6} $), wavevectors generically rotate towards the $ k^3 $-axis.

In the exceptional cases when $ \vec k $ is aligned to one of the principal axes, only the norm changes with time but the direction does not, as understood from \eqref{eq:k}.
Also, those quarter circumferences of the spheres connecting two principal axes (e.g.\ equator) are special in that if the endpoint of a wavevector lies on some of them, then it remains to do so, rotating from the direction of the faster axis to the slower.
This is understood as follows.
In order to parameterise a wavevector pointing somewhere on one of such quarter circumferences, we shall denote quantities associated with the axis of faster expansion by a superscript ``(fast)'' and those with the slower axis by ``(slow)''.
For example, if one considers a circumference specified by $ k_i = 0 $\,, then $ k^{(\mathrm{fast})}_i $ and $ k^{(\mathrm{slow})}_i $ denote the covariant components of the wavevector along the faster and the slower axes, respectively.
Then, we introduce an angle parameter $ \psi_i $ and a modulus $ k_0 $ as
\begin{equation}
\tan\psi_i
\equiv
  \frac{k^{(\mathrm{fast})}_i}{k^{(\mathrm{slow})}_i}\,,
\quad
k_0
\equiv
  \sqrt{\left(k^{(\mathrm{slow})}_i\right)^2 + \left(k^{(\mathrm{fast})}_i\right)^2}\,.
\label{eq:psik0}
\end{equation}
Hence, the ratio of the contravariant components evolves during the anisotropic phase, $ t \ll H^{-1} $\,, as
\begin{equation}
\begin{aligned}
\frac{k^{i(\mathrm{fast})}}{k^{i(\mathrm{slow})}}
&
= \tan\psi_i\,
  \left[\tanh\left(\frac{3Ht}{2}\right)\right]^{2 (q_i^{(\mathrm{slow})} - q_i^{(\mathrm{fast})})} \\
&
\simeq
  \tan\psi_i\,
  \left(\frac{3Ht}{2}\right)^{2 (q_i^{(\mathrm{slow})} - q_i^{(\mathrm{fast})})}\,.
\end{aligned}
\label{eq:k_dir}
\end{equation}

If $ k^{(\mathrm{fast})}_i < k^{(\mathrm{slow})}_i $\,, the wavevector is \emph{finally} aligned closer to the slower axis.
From eq.~\eqref{eq:k_dir}, the time at which $ \vec k $ comes to the midpoint of the two axes, i.e., at an angle of $ \frac{\pi}{4} $\,, is estimated as
\begin{equation}
H t_\mathrm{mid}
\simeq
  \frac{2}{3}\,
  \left(\frac{k^{(\mathrm{fast})}_i}{k^{(\mathrm{slow})}_i}\right)^{1/[2 (q_i^{(\mathrm{fast})} - q_i^{(\mathrm{slow})})]}\,.
\label{eq:t_mid}
\end{equation}
On the other hand, if $ k^{(\mathrm{fast})}_i > k^{(\mathrm{slow})}_i $\,, $ \vec k $ is initially closer to the faster axis and remains so until the end of the anisotropic Kasner regime.

From the above analysis, we can deduce that the direction of a wavevector pointing in general directions can only move within one of the quarter hemispheres bounded by such special quarter circumferences.
It follows that, thanks to the symmetry of the background, it is sufficient for our purpose to consider only one of those quarter hemispheres and we shall hereafter restrict on the one specified by $ k^i \geq 0 $ ($ i = 1,2,3 $).

\begin{figure}[htbp]
\begin{center}
\includegraphics[scale=0.5]{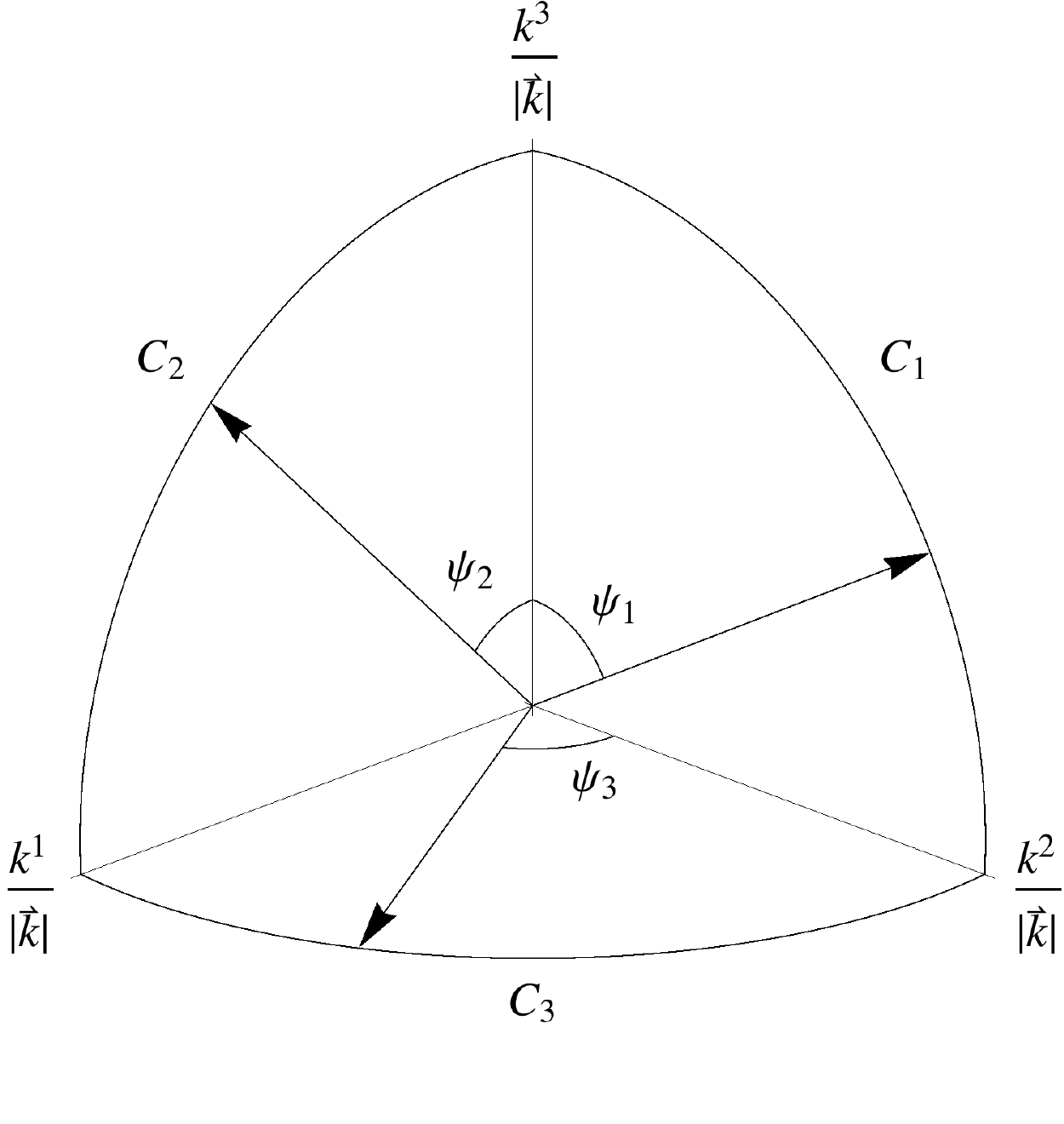}
\end{center}
\caption{\label{fig:circ}Illustration of the angles $ \psi_i $ and circumferences $ C_i $ on the unit sphere in the $ k $-space.
The arrows are examples of normalised final wavevector $ \lim_{t\to\infty} \vec{\hat k} = (k_1/k_0,k_2/k_0,k_3/k_0) $ lying on each $ C_i $\,.}
\end{figure}

In the $ k $-space, a normalised wavevector $ \vec{\hat k} = (k^1/|\vec k|,k^2/|\vec k|,k^3/|\vec k|) $ with one vanishing component $ k^i = 0 $ ($ i = 1,2,3 $) points on one of the quarter circumferences of the unit sphere, which we shall call $ C_i $\,.
See figure~\ref{fig:circ} for the definitions of the circumferences $ C_i $ and angles $ \psi_i $ ($ i = 1,2,3 $).

\subsection{\label{sec:axes}Evolution of modes aligned with the principal axes}

\begin{figure}[htbp]
\begin{center}
\includegraphics[scale=0.7]{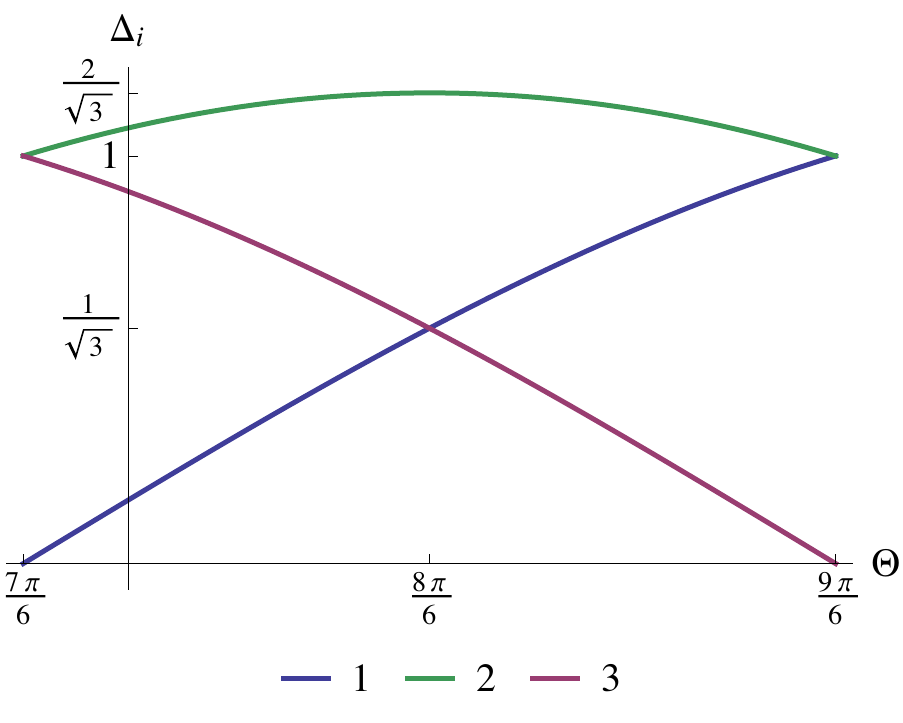}
\end{center}
\caption{\label{fig:Delta}Values of $ \Delta_i = \max_{j\neq i} q_j - \min_{j\neq i} q_j $\,.}
\end{figure}

We first consider the evolution of modes whose wavevector is aligned to either of the principal axes.
Let $ k_i $ be the only non-vanishing covariant component.
Then the wavevector $ \vec k $ is kept aligned with the $ k^i $-axis while its squared norm evolves with time as
\begin{equation}
k_i\,k^i
= k_i^2\,\tanh^{-2q_i}\left(\frac{3Ht}{2}\right)\,.
\label{eq:k_axes}
\end{equation}
The projected components of the shear tensor are greatly simplified in this case as\footnote{The sign of $ \sigma_+^{(\mathrm T)} $ is not needed because only its square will appear.}
\begin{equation}
\sigma^{(\mathrm S)}
= 3 q_i\,\frac{a_\mathrm{iso} H}{(a/a_\mathrm{iso})^2}\,,
\quad
\left|\sigma_+^{(\mathrm T)}\right|
= \frac{3 \Delta_i}{\sqrt 2}\,\frac{a_\mathrm{iso} H}{(a/a_\mathrm{iso})^2}\,,
\quad
\sigma_{(1)}^{(\mathrm V)}
= \sigma_{(2)}^{(\mathrm V)}
= \sigma_\times^{(\mathrm T)}
= 0\,,
\label{eq:shear_axes}
\end{equation}
where we have introduced $ \Delta_i \equiv \max_{j\neq i} q_j - \min_{j\neq i} q_j $\,, which quantifies the $ + $-component of the shear.
See figure~\ref{fig:Delta} for the dependence of $ \Delta_i $ on $ \Theta $ for each $ i $\,.
For $ \sigma_\times^{(\mathrm T)} = 0 $, the interaction term $ \xi $ in the gravitational-wave equations vanishes identically, see \eqref{eq:xi}, so the two polarisation modes are decoupled.

Substituting eqs.~\eqref{eq:k_axes} and \eqref{eq:shear_axes} into \eqref{eq:omega}, we obtain the squared frequencies $ \omega_\lambda^2 $ as
\begin{equation}
\begin{aligned}
\frac{\omega_\times^2}{a_\mathrm{iso}^2 H^2}
&
= \frac{k_i^2}{a_\mathrm{iso}^2 H^2}\,
  \tanh^{-2q_i}\left(\frac{3Ht}{2}\right)
  + \frac{2 - \cosh(6Ht)}{(a/a_\mathrm{iso})^4}
  - \frac{9 \Delta_i^2}{(a/a_\mathrm{iso})^4}\,, \\
\frac{\omega_+^2}{a_\mathrm{iso}^2 H^2}
&
= \frac{k_i^2}{a_\mathrm{iso}^2 H^2}\,
  \tanh^{-2q_i}\left(\frac{3Ht}{2}\right)
  + \frac{2 - \cosh(6Ht)}{(a/a_\mathrm{iso})^4}
  + \frac{6 \Delta_i^2\,(a/a_\mathrm{iso})^2}
         {\left[(2/3)\,\cosh(3Ht) - q_i\right]^2}\,.
\end{aligned}
\label{eq:omega_axes}
\end{equation}
The leading contributions from each term during the early anisotropic regime ($ t \ll t_\mathrm{iso} = H^{-1} $) are found as
\begin{equation}
\begin{aligned}
\frac{\omega_\times^2}{a_\mathrm{iso}^2 H^2}
&
\supset
  \frac{k_i^2}{a_\mathrm{iso}^2 H^2}\,
  \left(\frac{3Ht}{2}\right)^{-2q_i}
  + \frac{1}{(3Ht)^{4/3}}
  - \frac{9 \Delta_i^2}{(3Ht)^{4/3}}\,, \\
\frac{\omega_+^2}{a_\mathrm{iso}^2 H^2}
&
\supset
  \frac{k_i^2}{a_\mathrm{iso}^2 H^2}\,
  \left(\frac{3Ht}{2}\right)^{-2q_i}
  + \frac{1}{(3Ht)^{4/3}}
  + \frac{6 \Delta_i^2\,(3Ht)^{2/3}}{(2/3-q_i)^2}\,,
\end{aligned}
\label{eq:omega_axes_sim}
\end{equation}
where we have assumed $ q_i < \frac{2}{3} $\,.
The shear tensor gives a negative contribution $ \propto -\Delta_i^2/(Ht)^{4/3} $ to $ \omega_\times^2 $\,, which is missing in $ \omega_+^2 $ in contrast.

\begin{figure}[htbp]
\begin{center}
\includegraphics[scale=0.7]{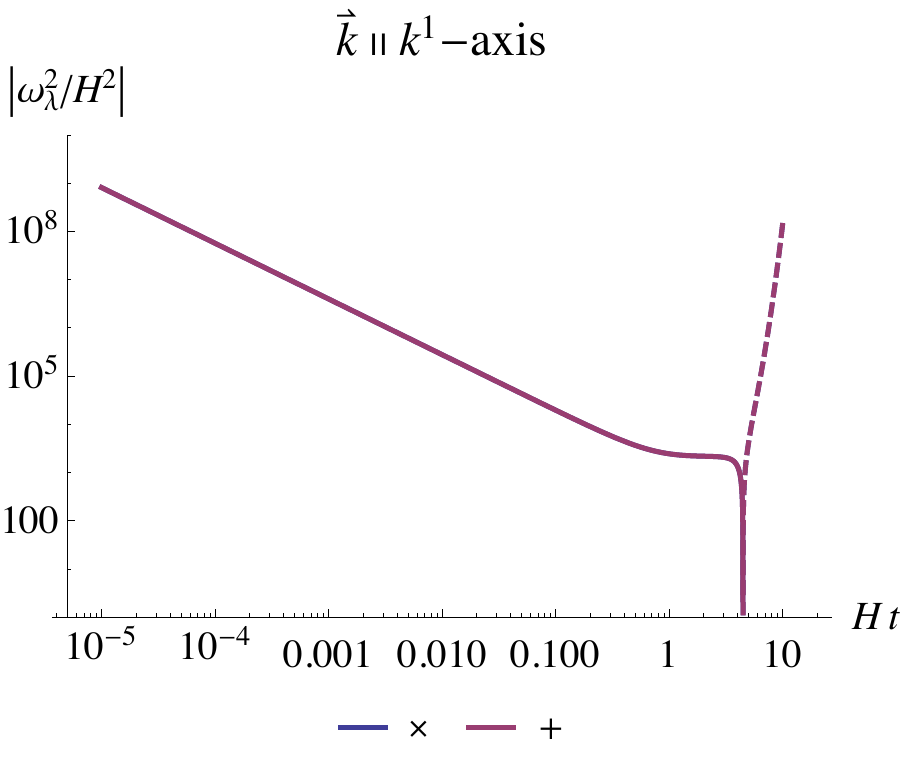}
\includegraphics[scale=0.7]{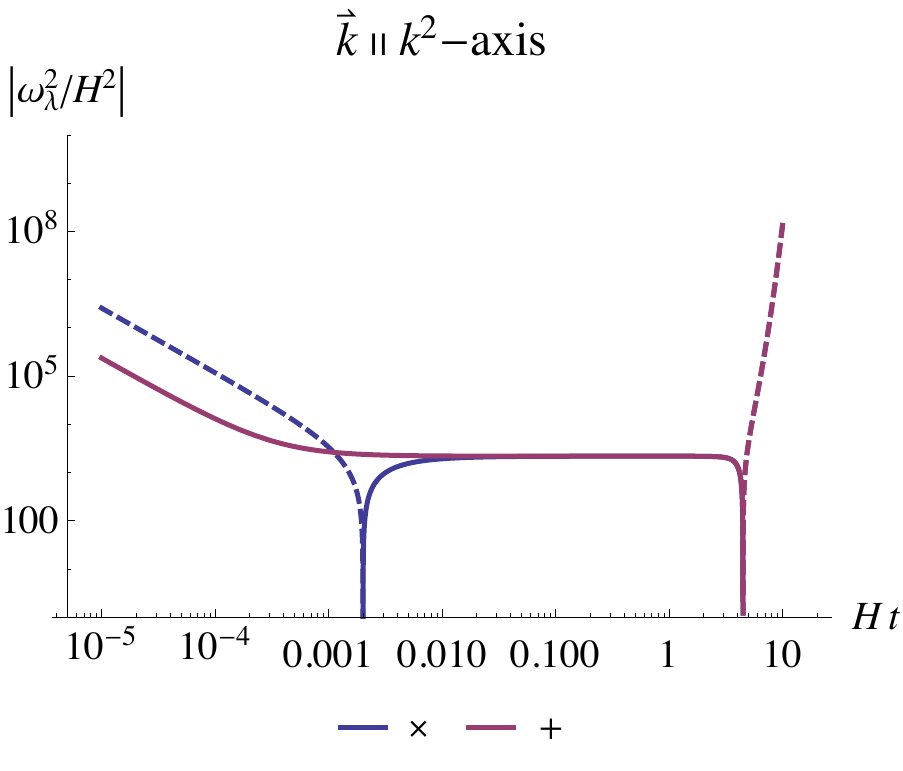} \\
\includegraphics[scale=0.7]{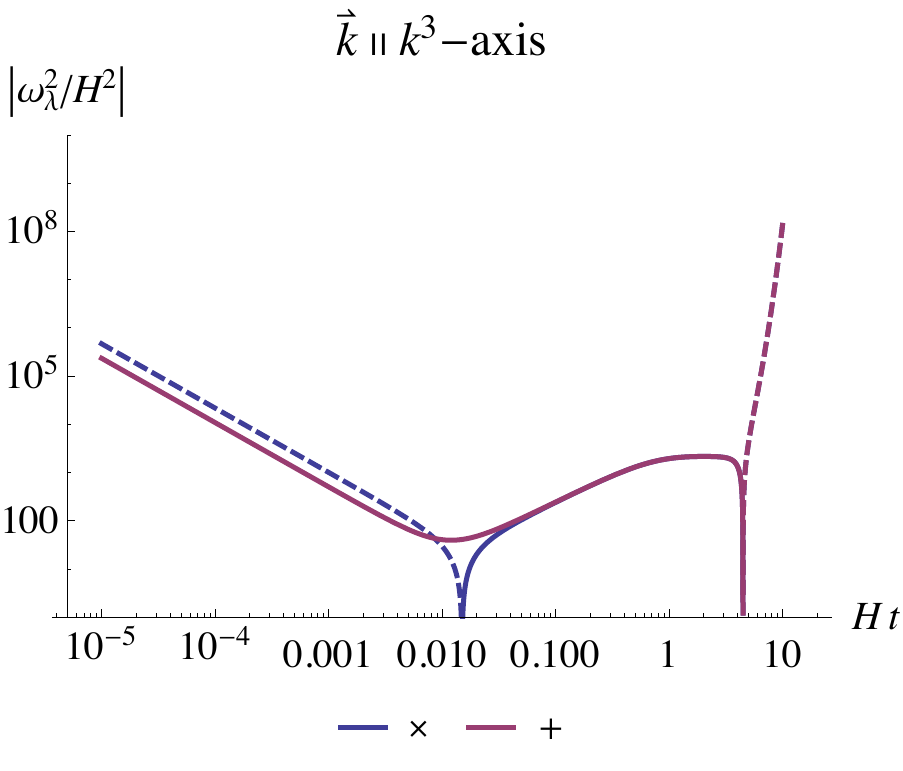}
\end{center}
\caption{\label{fig:omega_axes}Time evolutions of $ \omega_\times^2 $ (blue) and $ \omega_+^2 $ (red) for modes aligned with the $ k^1 $-axis (top-left), $ k^2 $-axis (top-right), and $ k^3 $-axis (bottom), for all of which the final wavenumber is $ k_i = 100\,a_\mathrm{iso}\,H $ ($ i = 1,2,3 $)\,.
The lines are dashed where $ \omega_\times^2 $ is negative.
The background anisotropy parameter is $ \Theta = \frac{8\pi}{6} $ corresponding to $ (q_1,q_2,q_3) = (\frac{1}{\sqrt 3},0,-\frac{1}{\sqrt 3}) $ and $ (\Delta_1,\Delta_2,\Delta_3) = (\frac{1}{\sqrt 3},\frac{2}{\sqrt 3},\frac{1}{\sqrt 3}) $.}
\end{figure}

In figure~\ref{fig:omega_axes}, we show typical time evolutions of $ \omega_\times^2 $ (blue) and $ \omega_+^2 $ (red) for modes aligned with either of the three principal axes.
The three figures correspond to the cases when $ \vec k \parallel k^1\textnormal{-axis} $ (top-left), $ \vec k \parallel k^2\textnormal{-axis} $ (top-right), and $ \vec k \parallel k^3\textnormal{-axis} $ (bottom), where the final wavenumber is $ k_i = 100\,a_\mathrm{iso}\,H $ ($ i = 1,2,3 $) and the background anisotropy parameter is $ \Theta = \frac{8\pi}{6} $\,, for which $ (q_1,q_2,q_3) = (\frac{1}{\sqrt 3},0,-\frac{1}{\sqrt 3}) $ and $ (\Delta_1,\Delta_2,\Delta_3) = (\frac{1}{\sqrt 3},\frac{2}{\sqrt 3},\frac{1}{\sqrt 3}) $\,.
Since $ \Delta_i > \frac{1}{3} $ for all $ i $ in this case, the squared frequency $ \omega_\times^2 $ should turn negative once the shear term dominates as implied by \eqref{eq:omega_axes_sim}.
To illuminate this, the lines for $ \omega_\times^2 $ are dashed when taking negative values.
It is observed that there is a large negative contribution to $ \omega_\times^2 $ from the shear at the earliest stage of the anisotropic phase ($ t \ll t_\mathrm{iso} = H^{-1} $) except for the $ k^1 $-axis-aligned mode (top-left panel), for which $ \omega_\times^2 $ is dominated by the $ k_i\,k^i $ term throughout the time range considered, $ 10^{-5} < H t \lesssim 1 $, although the shear would eventually dominate if one were allowed to go back in time indefinitely.

\begin{figure}[htbp]
\centering
\includegraphics[scale=0.7]{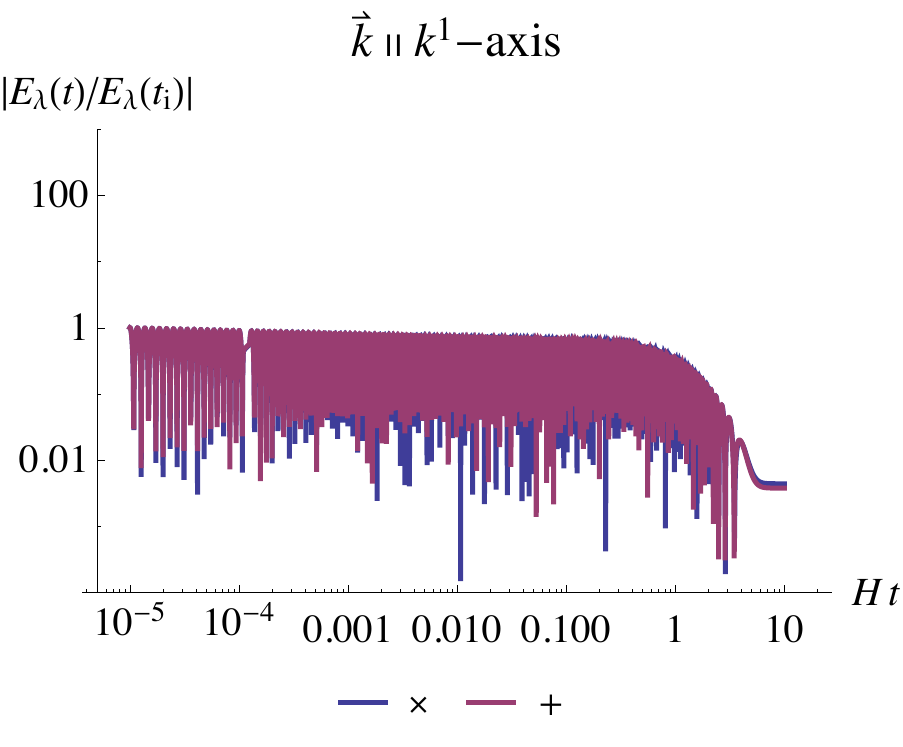}
\includegraphics[scale=0.7]{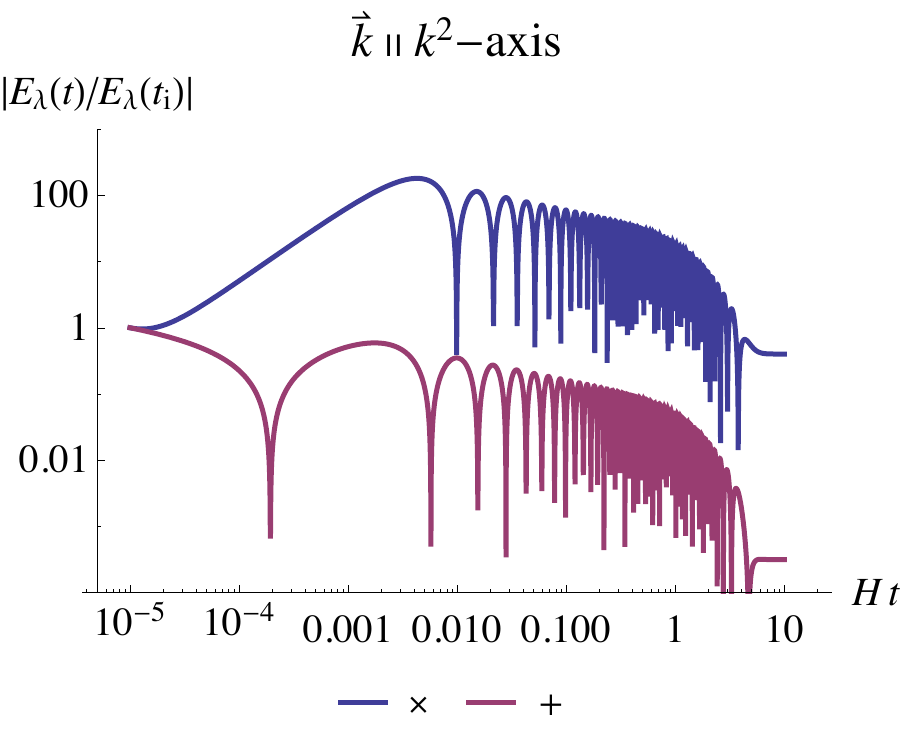} \\
\includegraphics[scale=0.7]{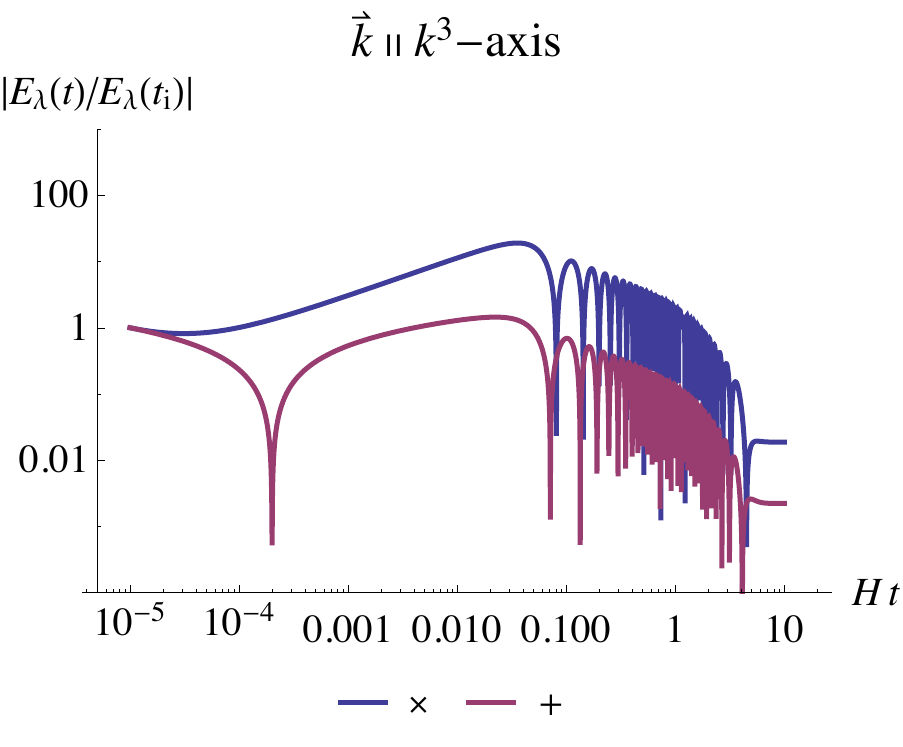}
\caption{\label{fig:E_axes}Waveforms of $ E_\times $ (blue) and $ E_+ $ (red) for the modes aligned with the $ k^1 $-axis (top-left), $ k^2 $-axis (top-right), and $ k^3 $-axis (bottom).
The initial time is $ t_\mathrm{ini} = 10^{-5}\,H^{-1} $ and the other parameters are the same as in figure~\ref{fig:omega_axes}.}
\end{figure}

Shown in figure~\ref{fig:E_axes} are the waveforms of the above three axis-aligned modes obtained by numerically integrating \eqref{eq:eom}.
As expected from the behaviours of $ \omega_\times^2 $\,, the $ \times $-mode grows substantially while the shear term is giving a dominant contribution.

These waveforms of gravitational waves can be analytically understood as follows.
First let us consider the $ \times $-mode.
The asymptotic form of $ \omega_\times^2 $ for $ H t \ll 1 $ is
\begin{equation}
\omega_\times^2
\sim
  \frac{k_i^2}{\left(2^{1/3} a_\mathrm{iso} H \eta\right)^{3q_i}}
  + \frac{1 - 9 \Delta_i^2}{4 \eta^2}\,,
\end{equation}
where, in converting the time coordinates, we have used a relation
\begin{equation}
H t
= \frac{2 \sqrt 2}{3}\,\left(a_\mathrm{iso} H \eta\right)^{3/2}\,
  \left[
   1
   + \frac{\left(a_\mathrm{iso} H \eta\right)^3}{6}
   + \mathcal O(a_\mathrm{iso} H \eta)^6
  \right]\,.
\end{equation}
Since $ q_i < \frac{2}{3} $ for a triaxial KdS metric, $ \omega_\times^2 $ is dominated by the shear term ($ \propto \eta^{-2} $) at earlier times and by the $ k_i\,k^i $ term ($ \propto \eta^{-3q_i} $) at later times.
The growth of $ \times $-mode is expected to stall around the transition between these two regimes, whose time is estimated as
\begin{equation}
a_\mathrm{iso} H \eta_\mathrm{stall}
= \left[
   \frac{\left|9 \Delta_i^2 - 1\right|}{2^{2-q_i}}\,
   \left(\frac{a_\mathrm{iso} H}{k_i}\right)^2
  \right]^{1/(2-3q_i)}
\end{equation}
or equivalently
\begin{equation}
H t_\mathrm{stall}
= \frac{2 \sqrt 2}{3}\,
  \left[
   \frac{\left|9 \Delta_i^2 - 1\right|}{2^{2-q_i}}\,
   \left(\frac{a_\mathrm{iso} H}{k_i}\right)^2
  \right]^{1/[2\,(2/3-q_i)]}\,.
\label{eq:t_stall}
\end{equation}

For $ \eta \ll \eta_\mathrm{stall} $\,, the equation of motion for the $ \times $-mode is approximated as
\begin{equation}
\mu_\times'' 
+ \frac{1 - 9 \Delta_i^2}{4 \eta^2}\,\mu_\times
\approx
  0
\label{eq:eom_axes}
\end{equation}
and its general solution is given by
\begin{equation}
\mu_\times
\approx
  C_+\,(H \eta)^{(1+3\Delta_i)/2}
  + C_-\,(H \eta)^{(1-3\Delta_i)/2}\,.
\end{equation}
Hence, up to the time $ \eta_\mathrm{stall} $\,, the growing mode behaves as $ |E_\times| \propto (H\eta)^{3\Delta_i/2} \propto (Ht)^{\Delta_i} $\,.

Once the shear term becomes subdominant after $ \eta_\mathrm{stall} $\,, the $ \times $-mode starts to oscillate obeying
\begin{equation}
\mu_\times''
+ \frac{k_i^2}{\left(2^{1/3} a_\mathrm{iso} H \eta\right)^{3q_i}}\,\mu_\times
\approx
  0\,.
\end{equation}
This cannot be integrated analytically for a general $ q_i $\,, but, since the exponents $ q_i $ only mildly depend on $ \Theta $\,, here we make a crude approximation that $ (q_1,q_2,q_3) \approx (\frac{2}{3},0,-\frac{2}{3}) $.
Then, we have a rough estimate
\begin{equation}
|\mu_\times|
\propto
\begin{cases}
\eta^{1/2}
& (i=1) \\
\eta^0
& (i=2) \\
\eta^{-1/2}
& (i=3)
\end{cases}
\end{equation}
and, therefore, we may estimate the amplitude of $ E_\times $ between $ t_\mathrm{stall} $ and $ t_\mathrm{iso} = H^{-1} $ as
\begin{equation}
|E_\times|
\propto
  (Ht)^{-p_i}
\quad\textnormal{with}\quad
(p_1,p_2,p_3)
\equiv
  \left(0,\frac{1}{3},\frac{2}{3}\right)\,.
\label{eq:E_oscc}
\end{equation}
As we will show numerically later, although quantitatively not quite precise, this approximation captures certain qualitative features of the evolution of gravitational waves during the Kasner epoch.

The evolution of the $ + $-mode can be deduced by setting $ \Delta_i = 0 $, see eq.~\eqref{eq:omega_axes_sim}.
Hence $ E_+ $ oscillates constantly until $ t_\mathrm{stall} $\,, and then decreasingly after $ t_\mathrm{stall} $ as
\begin{equation}
|E_+|
\propto
  (Ht)^{-p_i}\,.
\label{eq:E_oscp}
\end{equation}

The background is isotropised at $ \sim t_\mathrm{iso} $ and enters the standard de Sitter inflation phase.
During inflation, amplitudes of the both modes decay exponentially as $ E_\lambda \propto a^{-1} \propto \mathrm e^{-Ht} $ until the modes exit the Hubble radius when $ a(t) = k_i\,H^{-1} $\,, and are then frozen.
The amplitude at the end of inflation is therefore suppressed by a factor $ a(t_\mathrm{iso})\,H/k_i $ relative to the value at $ t = t_\mathrm{iso} $\,.

Summing up all these effects, the linear growth factors for each polarisation mode aligned with the principal axes are defined and estimated as
\begin{equation}
\begin{aligned}
D_\times
&
\equiv
  \lim_{t\to \infty}
  \left|\frac{E_\times(t)}{E_\times(t_\mathrm{ini})}\right|
\sim
  \frac{a_\mathrm{iso} H}{k_i}\,
  (H t_\mathrm{stall})^{p_i}\,
  \left(\frac{H t_\mathrm{stall}}{H t_\mathrm{ini}}\right)^{\Delta_i}\,, \\
D_+
&
\equiv
  \lim_{t\to \infty}
  \left|\frac{E_+(t)}{E_+(t_\mathrm{ini})}\right|
\sim
  \frac{a_\mathrm{iso} H}{k_i}\,
  (H t_\mathrm{stall})^{p_i}\,.
\end{aligned}
\label{eq:growth_axes}
\end{equation}

\begin{figure}[htbp]
\begin{center}
\includegraphics[scale=0.7]{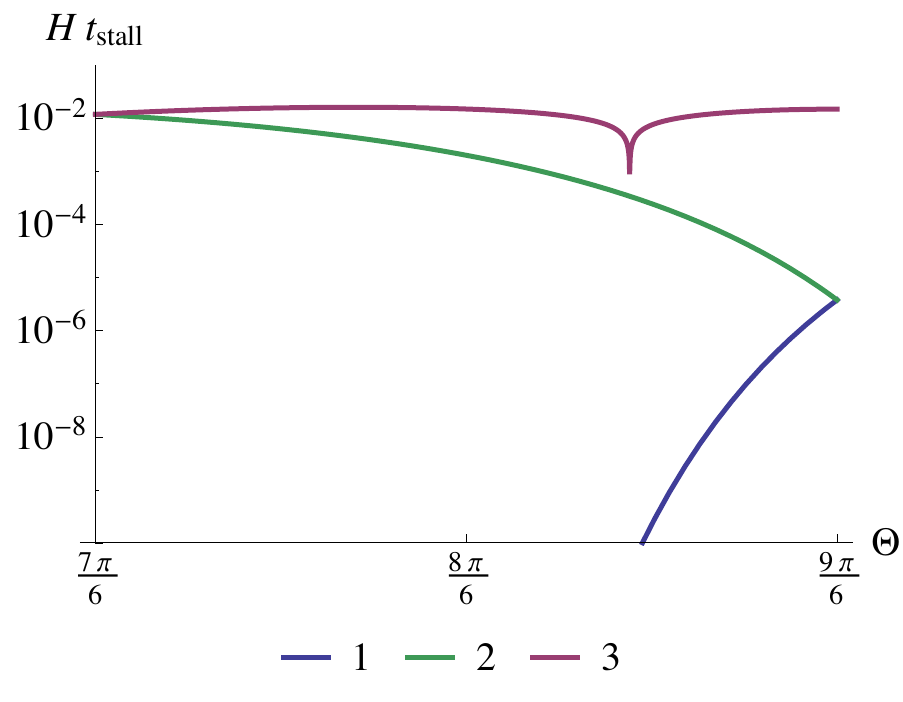}
\end{center}
\caption{\label{fig:t_stall}$ \Theta $-dependences of $ t_\mathrm{stall} $ for  modes aligned with each principal axis with the final wavenumber $ k_i = 100\,a_\mathrm{iso}\,H $ ($ i = 1,2,3 $).}
\end{figure}

Equation~\eqref{eq:t_stall} implies that smaller $ q_i $ leads to larger $ t_\mathrm{stall} $ as long as $ k_i \gg a_\mathrm{iso}\,H $\,.
Thus, the mode aligned with the $ k^3 $-axis enjoys the longest period of growth, see figure~\ref{fig:t_stall} for comparison of the values of $ t_\mathrm{stall} $ for the axis-aligned modes with the anisotropy parameter $ \Theta $ varied.
Note that $ H t_\mathrm{stall} $ given by \eqref{eq:t_stall} for $ i = 1 $ formally saturates to $ 1 $ near $ \Theta = \frac{7\pi}{6} $ (then $ q_\mathrm 1 \approx \frac{2}{3} $), but, since the usefulness of \eqref{eq:t_stall} is limited in such a situation, we have not shown the corresponding values for $ i = 1 $ in figure~\ref{fig:t_stall}.

\begin{figure}[htbp]
\begin{center}
\includegraphics[scale=0.7]{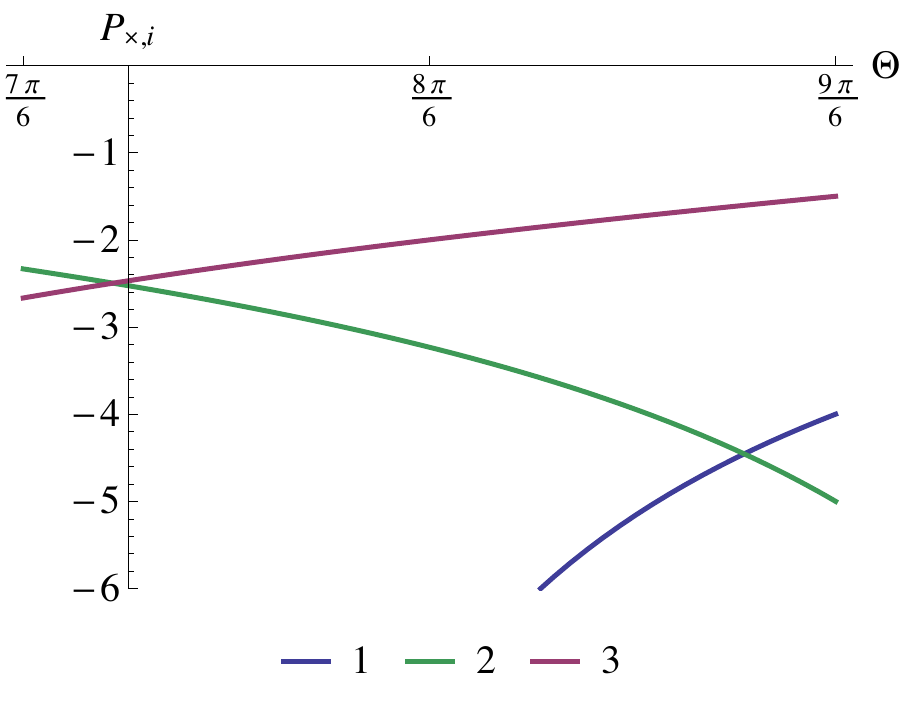}
\includegraphics[scale=0.7]{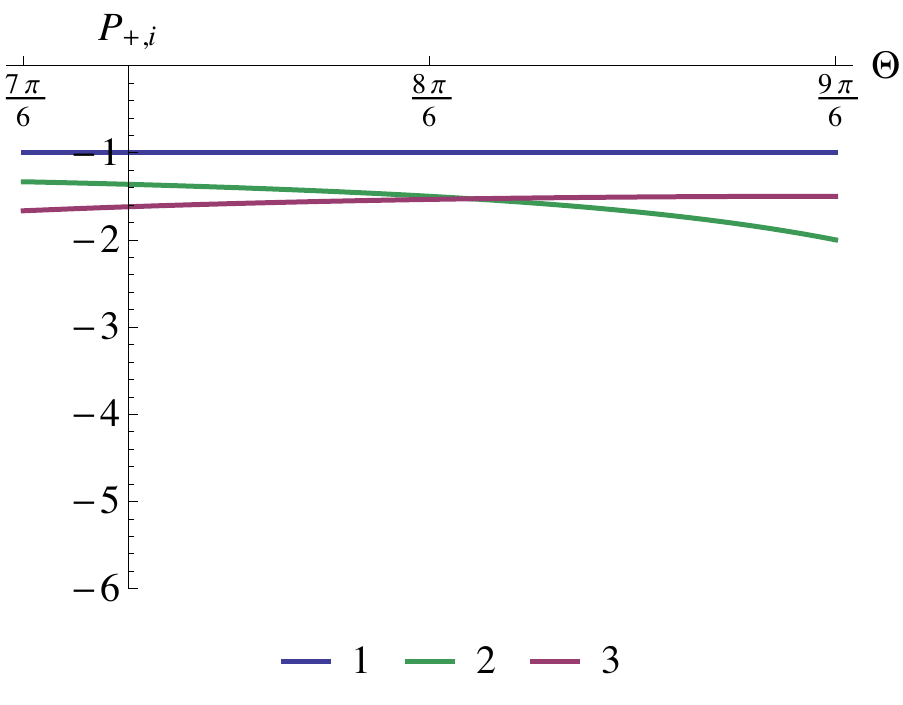}
\end{center}
\caption{\label{fig:index}$ \Theta $-dependences of the indices $ P_{\times,i} $ (left) and $ P_{+,i} $ (right).}
\end{figure}

Substituting eq.~\eqref{eq:t_stall} into \eqref{eq:growth_axes}, we can make the parameter dependence of the growth factors more explicit as
\begin{equation}
D_\times
\propto
  \left(\frac{k_i}{a_{\mathrm{iso}} H}\right)^{P_{\times,i}}\,
  (H t_\mathrm{ini})^{-\Delta_i}\,,
\quad
D_+
\propto
  \left(\frac{k_i}{a_{\mathrm{iso}} H}\right)^{P_{+,i}}
\label{eq:Dpi}
\end{equation}
with
\begin{equation}
P_{\times,i}
= -1 - \frac{p_i+\Delta_i}{2/3-q_i}\,,
\quad
P_{+,i}
= -1 - \frac{p_i}{2/3-q_i}\,.
\end{equation}
The validity of this expression for $ i = 1 $ is also degraded when $ \Theta \approx \frac{7\pi}{6} $ ($ q_i \approx \frac{2}{3} $) for the same reason as above.
In order to clarify the $ k_i $ dependences, we plot the indices $ P_{\times,i} $ and $ P_{+,i} $ in figure~\ref{fig:index}.
As for the $ \times $-mode, since $ P_{\times,3} $ is the largest of the three, the $ k^3 $-axis-aligned modes dominate at sufficiently large wavenumbers.

In the above analysis, we have implicitly assumed that the time scales are ordered as $ t_\mathrm{ini} < t_\mathrm{stall} < t_\mathrm{iso} $\,, but this does not necessarily hold, as in the case of the $ k^1 $-axis-aligned mode exemplified in this section.
If $ t_\mathrm{stall} < t_\mathrm{ini} $\,, both polarisation modes oscillate from the beginning with decaying amplitudes estimated by \eqref{eq:E_oscc} and \eqref{eq:E_oscp}, and the growth factor is estimated as
\begin{equation}
D_\times
\sim
  D_+
\sim
  \frac{a_\mathrm{iso} H}{k_i}\,
  (H t_\mathrm{ini})^{p_i}\,.
\label{eq:growth_axes2}
\end{equation}
On the other hand, although this is not likely to occur, if $ t_\mathrm{iso} < t_\mathrm{stall} $\,, the growth of gravitational waves does not stall until the time of isotropisation and the growth factor may be estimated as
\begin{equation}
D_\times
\sim
  D_+
\sim
  \frac{a_\mathrm{iso} H}{k_i}\,
  (H t_\mathrm{ini})^{-\Delta_i}\,.
\label{eq:growth_axes3}
\end{equation}

\begin{figure}[htbp]
\begin{center}
\includegraphics[scale=0.7]{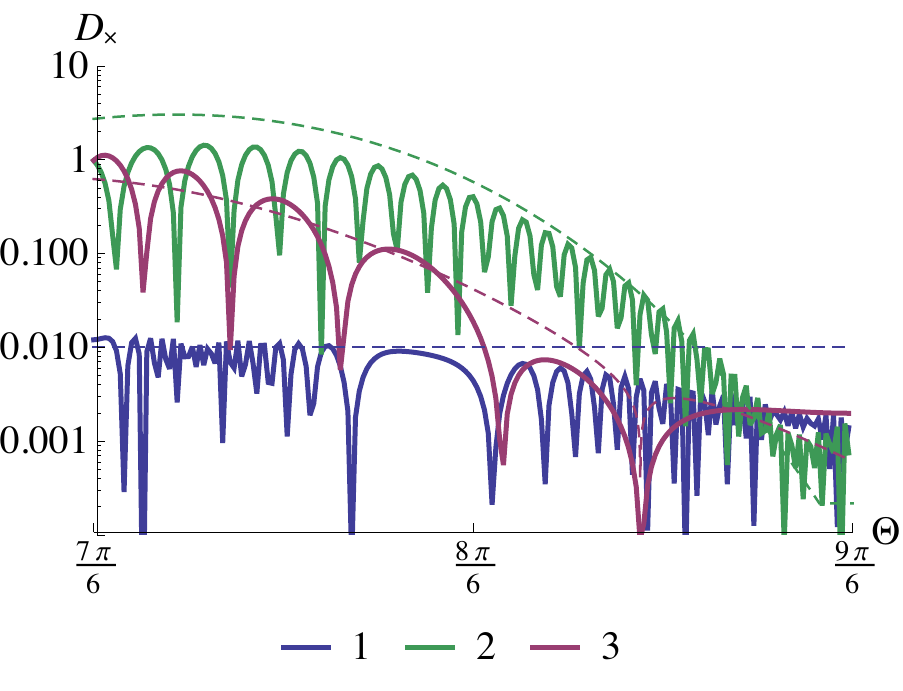}
\includegraphics[scale=0.7]{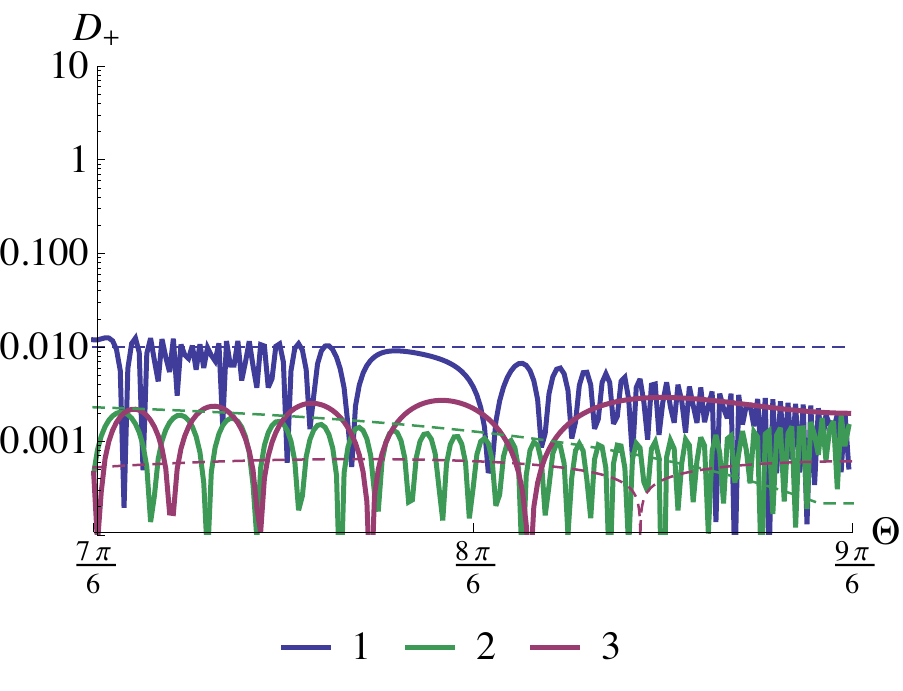}
\end{center}
\caption{\label{fig:D}$ \Theta $-dependences of the linear growth factors $ D_\times $ (left) and $ D_+ $ (right).
The final wavenumber is $ k_i = 100\,a_\mathrm{iso}\,H $ ($ i = 1,2,3 $) and the initial time is $ t_\mathrm{ini} = 10^{-5}\,H^{-1} $\,.
The dashed lines indicate the analytic estimate given by \eqref{eq:growth_axes}, in fairly good agreement with the numerical calculations (solid).}
\end{figure}

Figure~\ref{fig:D} shows the $ \Theta $-dependences of the growth factors for the axis-aligned modes with final wavenumber $ k_i = 100\,a_\mathrm{iso}\,H $ ($ i = 1,2,3 $) for the initial time $ t_\mathrm{ini} = 10^{-5}\,H^{-1} $\,.
The solid and dashed lines are the numerical and analytic estimate \eqref{eq:growth_axes}, respectively, which are in fairly good agreement with each other in the almost entire range of $ \Theta $\,.
The $ k^1 $-axis-aligned mode is almost independent of $ \Theta $ since it does not grow but constantly oscillates as implied by eqs.~\eqref{eq:E_oscc} and \eqref{eq:E_oscp}.
The $ k^2 $-axis-aligned and $ k^3 $-axis-aligned modes are comparable for the chosen set of parameters, although, as could be deduced from \eqref{eq:Dpi}, at sufficiently large wavenumbers the $ k^3 $-axis-aligned $ \times $-mode should dominate over the others.

\subsection{\label{sec:circs}Evolution of modes aligned between two principal axes}

Then, we extend our analysis to modes whose wavevector $ \vec k $ is not aligned with either of the principal axes but between two of them.
For such modes, as discussed in section~\ref{sec:k_dir}, the direction of a wavevector $ \vec k $ changes with time from the axis of faster expansion to the slower along one of the quarter circumferences $ C_i $ defined as in figure~\ref{fig:circ}.
As in section~\ref{sec:k_dir}, we introduce notations ``(fast)'' and ``(slow)'' for quantities associated with the faster and the slower axes, respectively.
Wavevectors to be considered should have one vanishing component, $ k_i = 0 $ ($ i = 1,2,3 $), while the other two, to be denoted as $ k_i^{(\mathrm{fast})} $ and $ k_i^{(\mathrm{slow})} $\,, are non-zero.
To parameterise such a wavevector, we introduce an angle parameter $ \psi_i $ and a modulus $ k_0 $ as given by eq.~\eqref{eq:psik0}.

It is notable that the $ \times $-component of the shear tensor $ \sigma_\times^{(\mathrm T)} $ still vanishes in this case, so, since $ \xi = 0 $, the two polarisation modes of gravitational waves are decoupled.
Therefore the evolution of each mode can still be analysed separately.

As we will discuss shortly, the distribution of the intensity of gravitational waves on the circumferences $ C_i $ will serve as a key clue to establish the connection between the directional variation of gravitational-wave intensity and the pre-inflationary parameters.
The situation on a circumference may be classified into the following three cases according to the final direction of the wavevector, i.e., the value of $ \tan\psi_i $\,.

First, if the final direction of a mode satisfies $ \tan\psi_i > 1 $, the wavevector $ \vec k $ remains close to the faster axis throughout the anisotropic Kasner regime.
In terms of $ t_\mathrm{mid} $ given by \eqref{eq:t_mid}, the condition for this to realise may be written as
\begin{equation}
t_\mathrm{mid}(\psi_i)
> t_\mathrm{iso}
\quad\Leftrightarrow\quad
\tan\psi_i
> 1\,.
\end{equation}
We will approximately evaluate the evolution of the modes of this kind by regarding their wavevectors as exactly aligned with the faster axis throughout the anisotropic phase.

Next, for modes with $ \tan\psi_i < 1 $, a turnover of the direction could happen at the time $ t = t_\mathrm{mid} $ given by eq.~\eqref{eq:t_mid}.
However, if
\begin{equation}
t_\mathrm{mid}(\psi_i)
< t_\mathrm{ini}
\quad\Leftrightarrow\quad
\tan\psi_i
< \left(\frac{3 H t_\mathrm{ini}}{2}\right)^{2 (q_i^{(\mathrm{fast})}-q_i^{(\mathrm{slow})})}\,,
\end{equation}
the wavevector $ \vec k $ is already close to the slower axis at the initial time $ t = t_\mathrm{ini} $\,, hence no turnover occurs during the Kasner epoch.
In this case the wavevector is regarded as aligned exactly with the slower axis throughout.

Finally, a turnover actually happens during the anisotropic phase if
\begin{equation}
t_\mathrm{ini}
< t_\mathrm{mid}(\psi_i)
< t_\mathrm{iso}
\quad\Leftrightarrow\quad
\left(\frac{3 H t_\mathrm{ini}}{2}\right)^{2 (q_i^{(\mathrm{fast})}-q_i^{(\mathrm{slow})})}
< \tan\psi_i
< 1\,.
\end{equation}
We regard such modes as aligned with the faster axis before $ t_\mathrm{mid} $ and with the slower axis after $ t_\mathrm{mid} $\,.
The analyses in the first and second cases above still apply to the time ranges $ t < t_\mathrm{mid} $ and $ t > t_\mathrm{mid} $\,, respectively.

\subsubsection{\label{sec:fast}Case 1: $ t_\mathrm{mid}(\psi_i) > t_\mathrm{iso} $}

For such a mode, we approximate the growth rate by the one for a mode exactly aligned with the faster axis and having a wavenumber $ k_i = k_i^{(\mathrm{fast})} = k_0 \sin\psi_i $\,.
Applying \eqref{eq:growth_axes} or \eqref{eq:growth_axes2} with $ q_i = q_i^{(\mathrm{fast})} $\,, $ \Delta_i = \Delta_i^{(\mathrm{fast})} $\,, and $ p_i = p_i^{(\mathrm{fast})} $\,, we may estimate the growth factor for the $ \times $-mode as
\begin{equation}
D_\times(\psi_i)
\sim
  \frac{a_\mathrm{iso} H}{k_0}
  \times
\begin{cases}
\left(H t^{(\mathrm{fast})}_\mathrm{stall}\right)^{p_i^{(\mathrm{fast})}}\,
\left(\frac{H t^{(\mathrm{fast})}_\mathrm{stall}}{H t_\mathrm{ini}}\right)^{\Delta_i^{(\mathrm{fast})}}
& (t_\mathrm{ini} < t_\mathrm{stall}^{(\mathrm{fast})}) \\
\left(H t_\mathrm{ini}\right)^{p_i^{(\mathrm{fast})}}\,
& (t_\mathrm{stall}^{(\mathrm{fast})} < t_\mathrm{ini})\,,
\end{cases}
\label{eq:growth_fast}
\end{equation}
where
\begin{equation}
H t^{(\mathrm{fast})}_\mathrm{stall}(\psi_i)
= \frac{2 \sqrt 2}{3}\,
  \left[
   \frac{\left|9 (\Delta_i^{(\mathrm{fast})})^2 - 1\right|}{2^{2-q_i^{(\mathrm{fast})}}}\,
   \left(\frac{a_\mathrm{iso} H}{k_0 \sin\psi_i}\right)^2
  \right]^{1/[2\,(2/3-q_i^{(\mathrm{fast})})]}\,.
\label{eq:t_stall_rapid}
\end{equation}

It is noted that, for both the circumferences $ C_2 $ and $ C_3 $\,, the $ k^1 $-axis serves as the axis of faster expansion, so $ q_i^{(\mathrm{fast})} = q_1 \geq \frac{1}{3} $ for $ i = 2,3 $\,.
Then, since $ k_0 > a_\mathrm{iso}\,H $ by assumption, a rough but strict upper limit on the time scale of the growth stall is given as
\begin{equation}
H t^{(\mathrm{fast})}_\mathrm{stall}(\psi_i)
\lesssim
  \left(\frac{k_0}{a_\mathrm{iso} H}\right)^{-3}
\quad
(i = 2,3)\,.
\end{equation}
This implies that growth of gravitational waves does not take place near the $ k^1 $-axis if the wavelength is sufficiently short such that $ k_0 \gg a_\mathrm{iso}\,H $\,.

\subsubsection{\label{sec:slow}Case 2: $ t_\mathrm{mid}(\psi_i) < t_\mathrm{ini} $}

Again, no turnover takes place.
Similarly to Case 1 but with the use of $ k_i = k_i^{(\mathrm{slow})} = k_0\,\cos\psi_i $ instead of $ k_0\,\sin\psi_i $\,, the growth factor is estimated as
\begin{equation}
D_\times(\psi_i)
\sim
  \frac{a_\mathrm{iso} H}{k_0}
  \times
\begin{cases}
\left(Ht_\mathrm{stall}^{(\mathrm{slow})}\right)^{p_i^{(\mathrm{slow})}}\,
\left(\frac{H t_\mathrm{stall}^{(\mathrm{slow})}}{H t_\mathrm{ini}}\right)^{\Delta_i^{(\mathrm{slow})}}
& (t_\mathrm{ini} < t_\mathrm{stall}^{(\mathrm{slow})}) \\
\left(Ht_\mathrm{ini}\right)^{p_i^{(\mathrm{slow})}}
& (t_\mathrm{stall}^{(\mathrm{slow})} < t_\mathrm{ini})
\end{cases}
\label{eq:growth_slow}
\end{equation}
with
\begin{equation}
H t^{(\mathrm{slow})}_\mathrm{stall}(\psi_i)
\approx
  \frac{2 \sqrt 2}{3}\,
  \left[
   \frac{\left|9 (\Delta_i^{(\mathrm{slow})})^2 - 1\right|}{2^{2-q_i^{(\mathrm{slow})}}}\,
   \left(\frac{a_\mathrm{iso} H}{k_0 \cos\psi_i}\right)^2
  \right]^{1/[2\,(2/3-q_i^{(\mathrm{slow})})]}\,.
\label{eq:t_stall_slow}
\end{equation}
In contrast to the previous case, the exponent for this expression does not become large except on $ C_3 $ for a nearly axisymmetric background with $ \Theta \approx \frac{9\pi}{6} $\,.

\subsubsection{Case 3: $ t_\mathrm{ini} < t_\mathrm{mid}(\psi_i) < t_\mathrm{iso} $}

Finally, we consider the intermediate case in which a turnover occurs during the Kasner regime.
We regard the wavevector $ \vec k $ as aligned with the faster axis before $ t_\mathrm{mid} $ and with the slower axis after $ t_\mathrm{mid} $\,.

\begin{figure}[htbp]
\begin{center}
\includegraphics[scale=0.7]{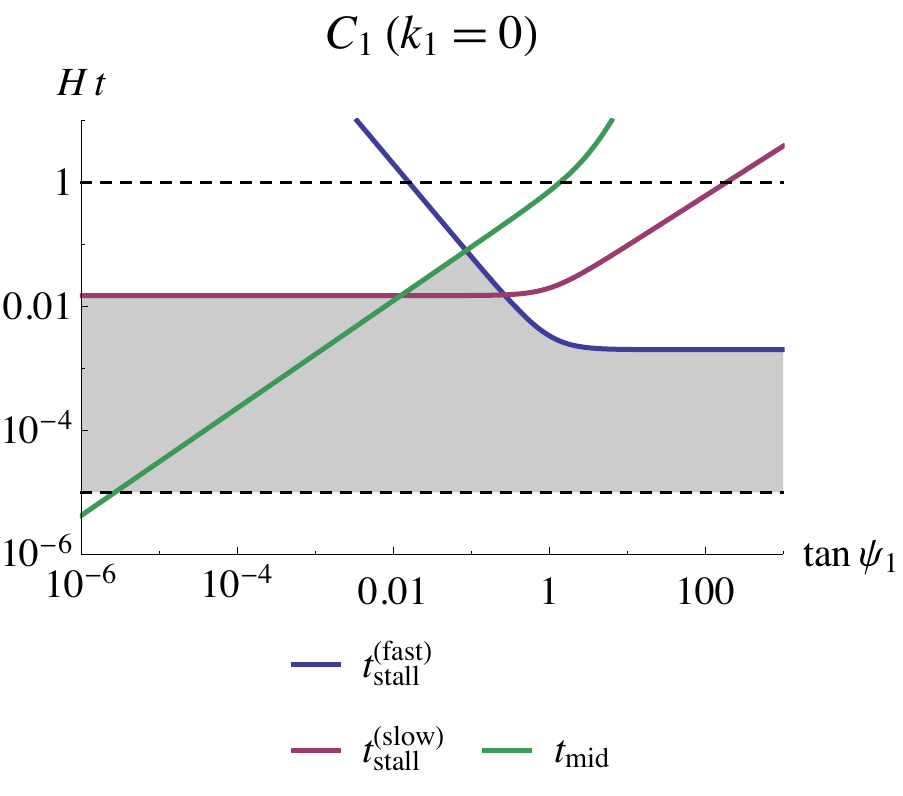}
\includegraphics[scale=0.7]{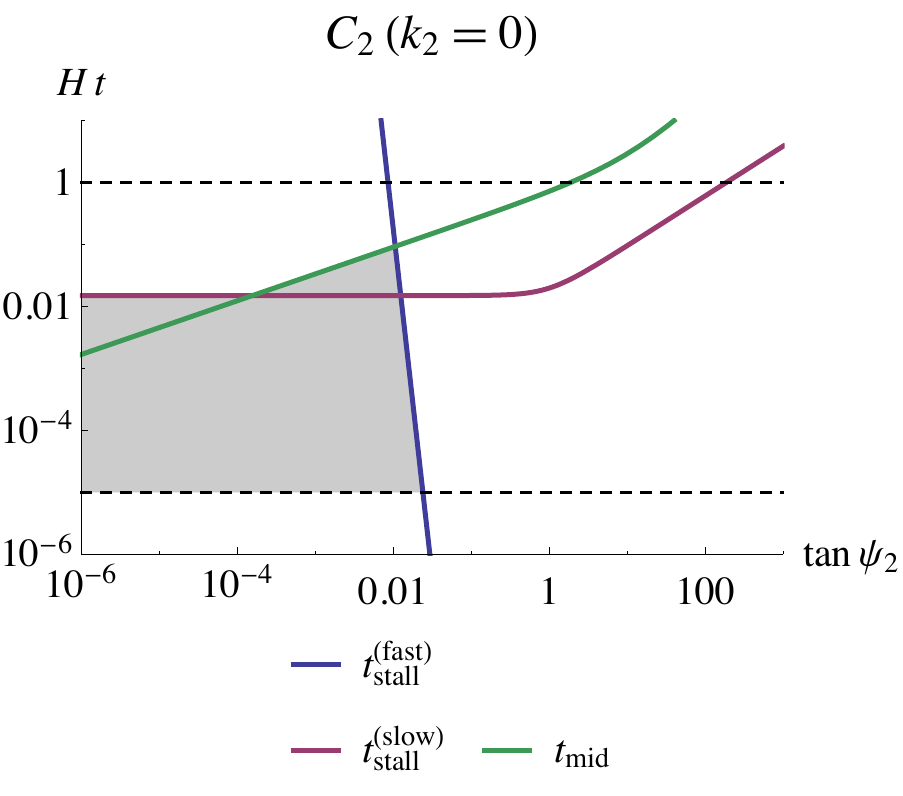} \\
\includegraphics[scale=0.7]{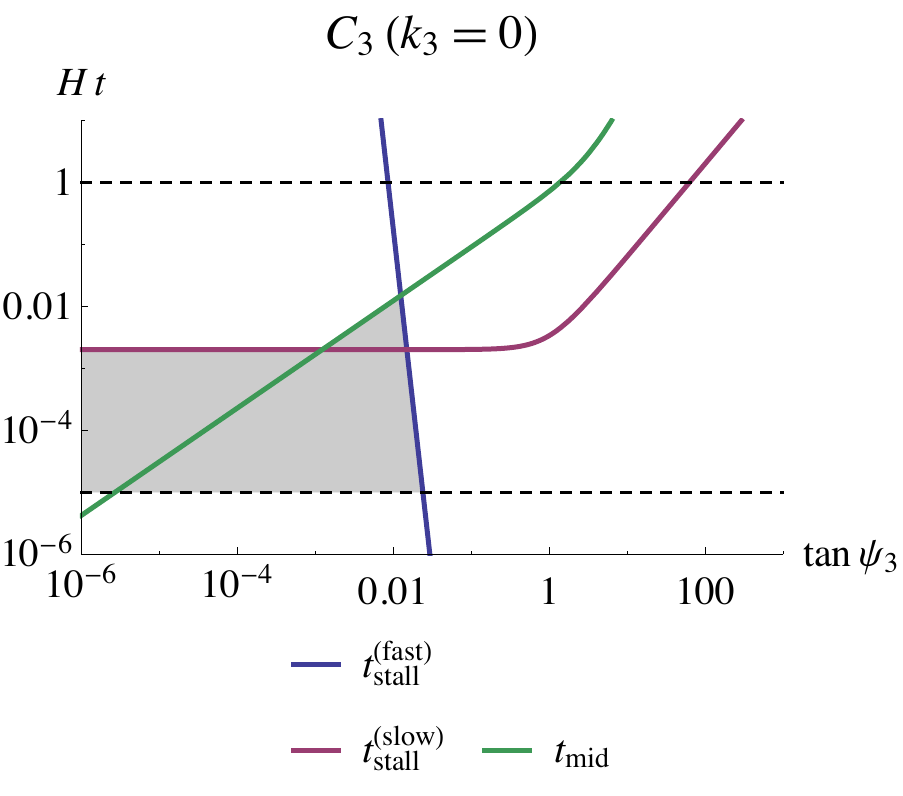}
\end{center}
\caption{\label{fig:times}Time scales on the circumferences $ C_1 $ (top-left), $ C_2 $ (top-right), and $ C_3 $ (bottom) for $ k_0 = 100\,a_\mathrm{iso}\,H $ and $ \Theta = \frac{8\pi}{6} $\,.
The horizontal dashed lines correspond to $ H t_\mathrm{iso} = 1 $ and $ H t_\mathrm{ini} = 10^{-5} $\,.
For each value of $ \tan\psi_i $\,, the mode can grow within the time range indicated in cray, see text.}
\end{figure}

It is useful here to compare the three time scales $ t_\mathrm{stall}^{(\mathrm{fast})} $\,, $ t_\mathrm{stall}^{(\mathrm{slow})} $\,, and $ t_\mathrm{mid} $\,.
Shown in figure~\ref{fig:times} are their values evaluated on the circumferences $ C_1 $ (top-left), $ C_2 $ (top-right), and $ C_3 $ (bottom) for the parameters $ k_0 = 100\,a_\mathrm{iso}\,H $ and $ \Theta = \frac{8\pi}{6} $\,.
As we will discuss shortly, the $ \times $-mode can grow during the time range indicated as grey-shaded regions.

Before the time of turnover $ t_\mathrm{mid} $ is reached, i.e., below the green lines in figure~\ref{fig:times}, the wavevector $ \vec k $ is considered to be aligned with the faster axis and the analysis in section~\ref{sec:fast} may be applied.
The growth stalls at $ t_\mathrm{stall}^{(\mathrm{fast})} $ if it comes before $ t_\mathrm{mid} $\,, but otherwise continues up to $ t_\mathrm{mid} $\,.
Indeed, in figure~\ref{fig:times}, there are small intervals of $ \tan\psi_i $ in which $ t_\mathrm{mid} < t_\mathrm{stall}^{(\mathrm{fast})} $\,.

After $ t_\mathrm{mid} $\,, i.e., above the green lines in figure~\ref{fig:times}, the wavevector $ \vec k $ is considered to be aligned with the slower axis and the analysis in section~\ref{sec:slow} may be applied.
If $ t_\mathrm{mid} < t_\mathrm{stall}^{(\mathrm{slow})} $\,, then the mode continues (or resumes) to grow beyond $ t_\mathrm{mid} $ up to $ t_\mathrm{stall}^{(\mathrm{slow})} $\,.

Near the $ k^1 $-axis on the circumferences $ C_2 $ and $ C_3 $\,, it may happen that the time $ t_\mathrm{stall}^{(\mathrm{fast})} $ comes to earlier than $ t_\mathrm{ini} $ and then growth is substantially suppressed.
Indeed, for our choice of parameters, the values of $ t_\mathrm{stall}^{(\mathrm{fast})} $ evaluated on $ C_2 $ and $ C_3 $ are much earlier than $ t_\mathrm{ini} = 10^{-5}\,H^{-1} $ as in the top-right and the bottom panels of figure~\ref{fig:times}.
If this is the case, a growth can only happen near the slower axes on these circumferences.
Therefore, a criterion for growth in terms of the angle from the slower axis may be given by 
\begin{equation}
t_\mathrm{stall}^{(\mathrm{fast})}
\gtrsim
  t_\mathrm{ini}
\quad\Leftrightarrow\quad
\psi_i
\lesssim
  \left(\frac{k_0}{a_\mathrm{iso} H}\right)^{-1}\,
  \left(H t_\mathrm{ini}\right)^{-(2/3-q_i^{(\mathrm{fast})})}
\quad
(i = 2,3)\,.
\label{eq:psi_th}
\end{equation}

In contrast, for the current choice of parameters, such suppression is not working on $ C_1 $ as understood from the top-left panel of figure~\ref{fig:times}, where $ t_\mathrm{stall}^{(\mathrm{fast})}(\tan\psi_1 \gtrsim 1) $ is comparable to $ t_\mathrm{stall}^{(\mathrm{slow})}(\tan\psi_1 \lesssim 1) $ and never comes below $ t_\mathrm{ini} $\,.
However, for some parameters and wavelengths it may happen that $ t_\mathrm{stall}^{(\mathrm{fast})}$ on $ C_1 $ becomes earlier than $ t_\mathrm{ini} $\,, leading to suppression of growth near the $ k^2 $-axis (the faster axis on $ C_1 $).
In the next sections, we will pay particular attention to this possibility.

\subsubsection{Growth factors}

\begin{figure}[htbp]
\begin{center}
\includegraphics[scale=0.7]{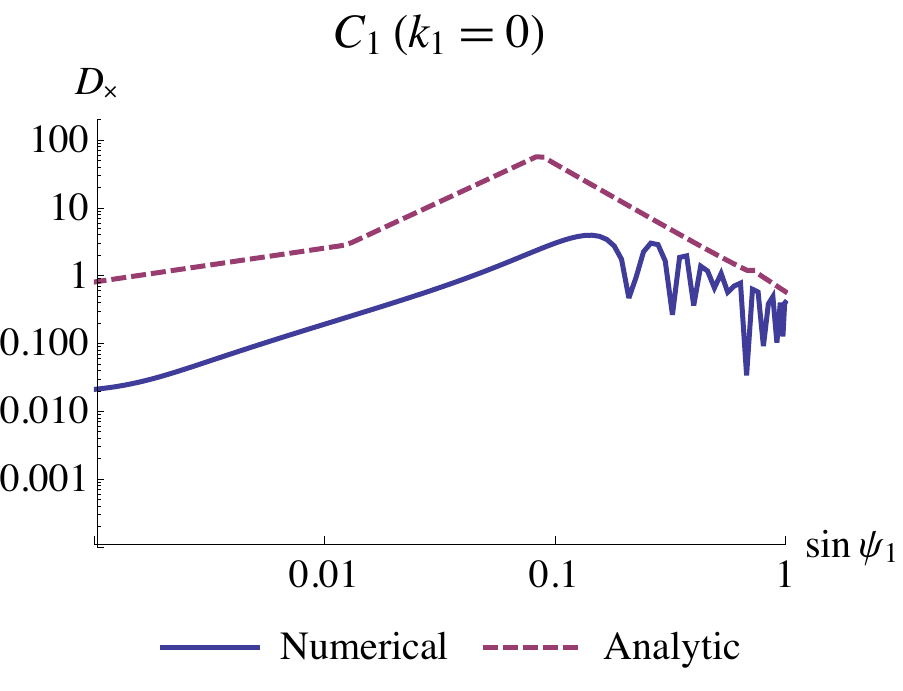}
\includegraphics[scale=0.7]{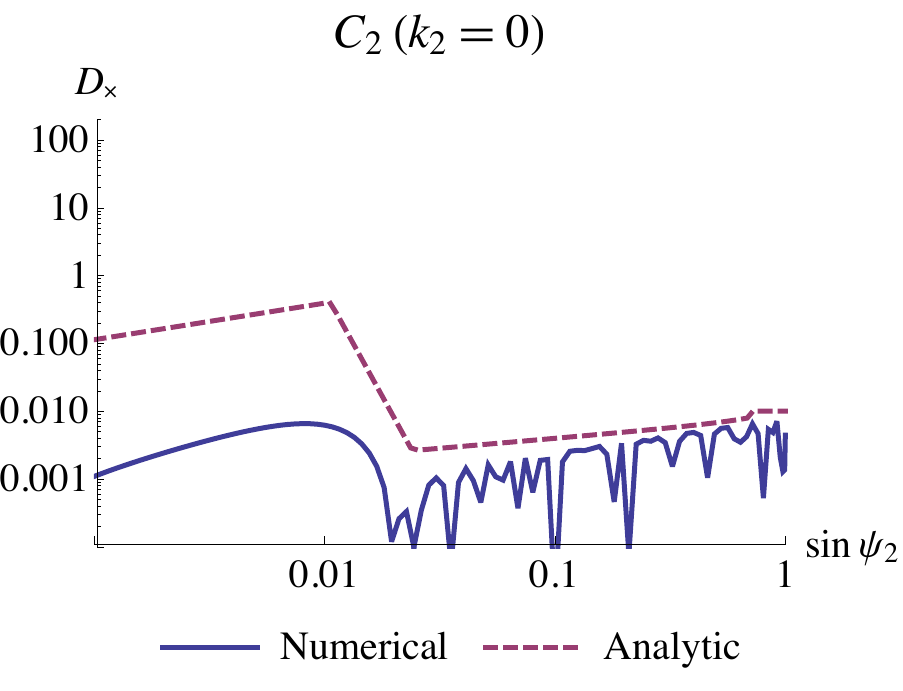} \\
\includegraphics[scale=0.7]{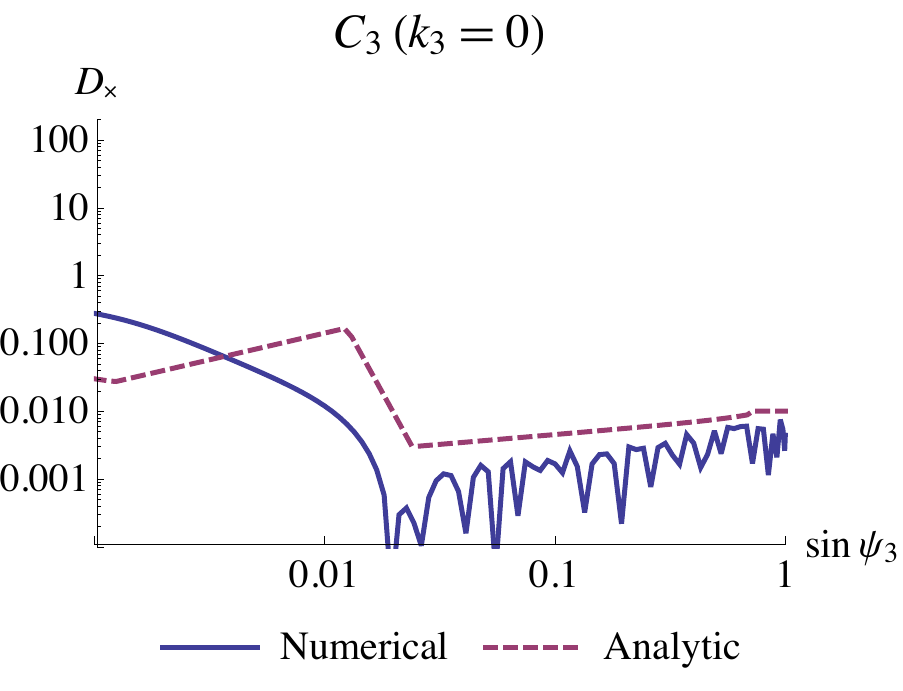}
\end{center}
\caption{\label{fig:Dc_circ}Linear growth factors for the $ \times $-modes on the circumferences $ C_1 $ (top-left), $ C_2 $ (top-right), and $ C_3 $ (bottom).
The background anisotropy parameter is $ \Theta = \frac{8\pi}{6} $\,.
The initial time and the final wavenumber are $ t_\mathrm{ini} = 10^{-5}\,H^{-1} $ and $ k_0 = 100\,a_\mathrm{iso}\,H $\,, respectively.}
\end{figure}

In figure~\ref{fig:Dc_circ}, we show the linear growth factors for the $ \times $-mode on the circumferences $ C_1 $ (top-left), $ C_2 $(top-right), and $ C_3 $ (bottom) for the parameters $ \Theta = \frac{8\pi}{6} $\,, $ t_\mathrm{ini} = 10^{-5}\,H^{-1} $\,, and $ k_0 = 100\,a_\mathrm{iso}\,H $\,.
In each figure, $ \psi_i $ measures the angle from the axis of slower expansion.
The analytic estimates based on the arguments in the previous sections are also plotted as dashed lines.

As understood from the figure, there is only weak contrast in growths on $ C_1 $\,, whereas rather sharp declines at some angle of $ \mathcal O(0.01) $ from the slower axes on $ C_2 $ and $ C_3 $ are indicated.
This implies that the region of high-intensity gravitational waves on the celestial sphere would form a belt-like pattern around a great circle containing $ C_1 $\,, which we will numerically confirm in the next section.

\begin{figure}[htbp]
\begin{center}
\includegraphics[scale=0.7]{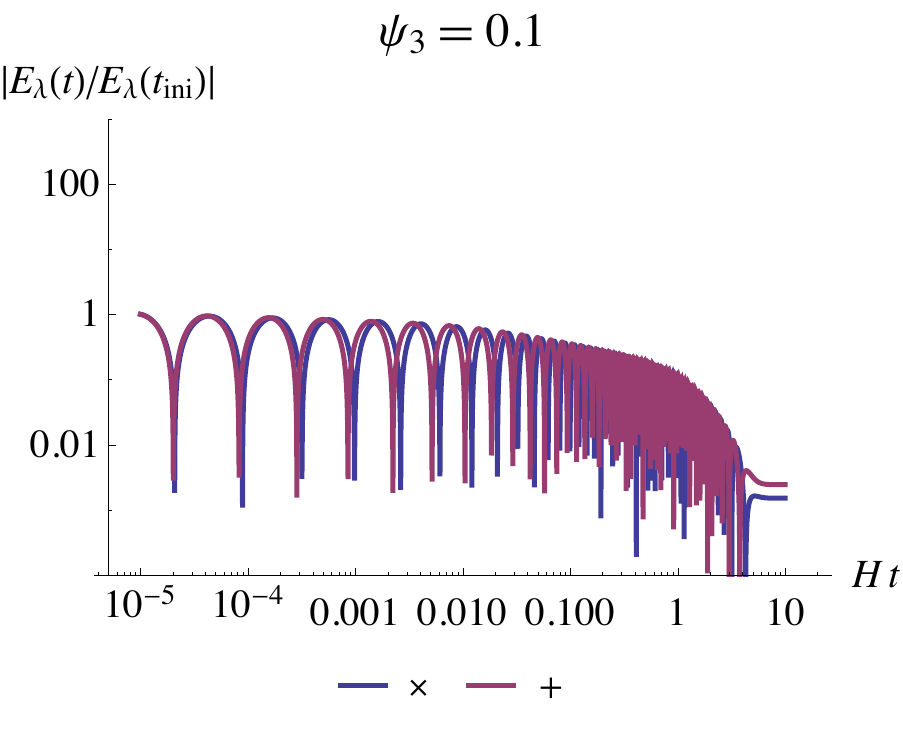}
\includegraphics[scale=0.7]{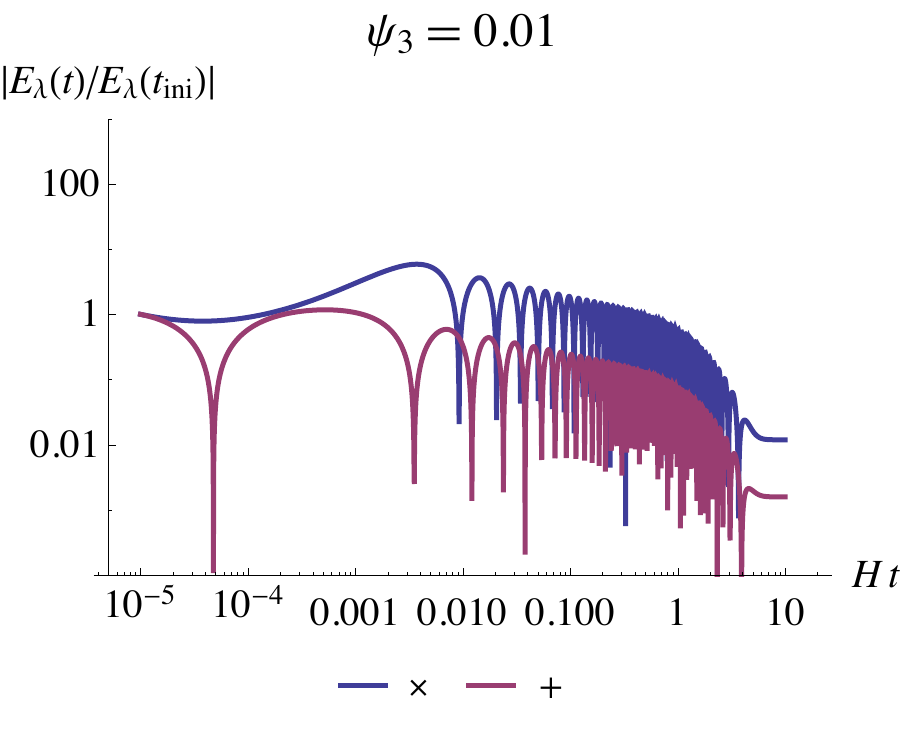} \\
\includegraphics[scale=0.7]{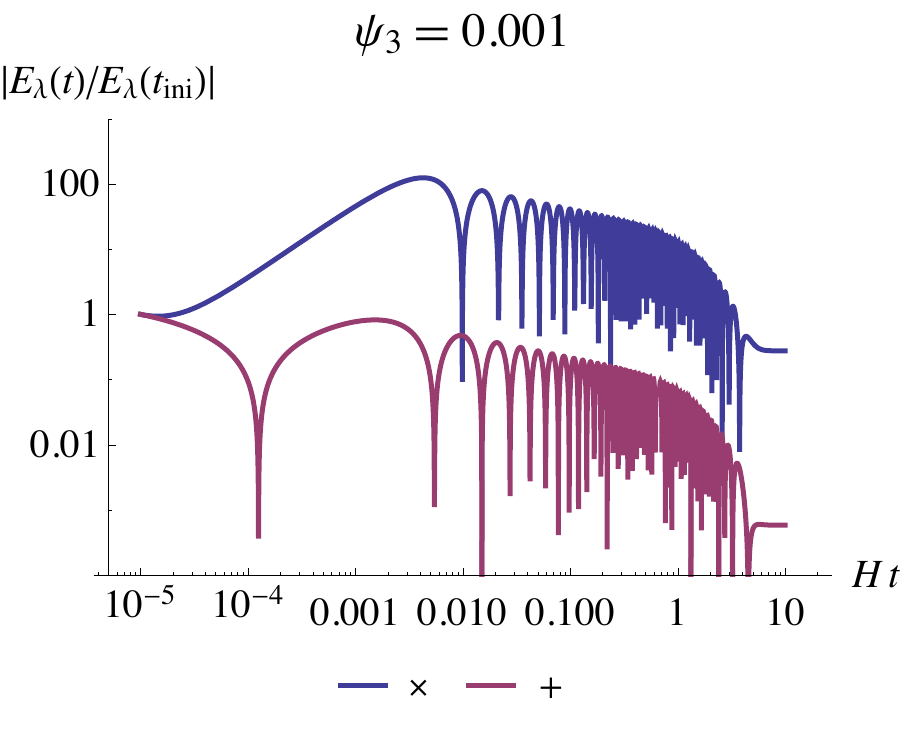}
\end{center}
\caption{\label{fig:E_C3}Waveforms of several modes on the circumference $ C_3 $ ($ k_3 = 0 $) with a common final wavenumber $ k_0 = 100\,a_\mathrm{iso}\,H $ but with different final directions $ k_1/k_2 = \tan\psi_3 $ for $ \psi_3 = 10^{-1} $ (top-left), $ 10^{-2} $ (top-right), and $ 10^{-3} $ (bottom).
The background anisotropy parameter and the initial time are $ \Theta = \frac{8\pi}{6} $ and $ t_\mathrm{ini} = 10^{-5}\,H^{-1} $\,, respectively.}
\end{figure}

In figure~\ref{fig:E_C3}, we show waveforms of $ E_\times $ (blue) and $ E_+ $ (red) for several modes on $ C_3 $\,, i.e., with $ k_3 = 0 $.
The modes have a common final wavenumber $ k_0 = \sqrt{k_1^2+k_2^2} = 100\,a_\mathrm{iso}\,H $ but different final directions parameterised by $ \tan\psi_3 = k_1/k_2 $\,: $ \psi_3 = 10^{-1} $ (top-left), $ 10^{-2} $ (top-right) and $ 10^{-3} $ (bottom).
The other parameters are the same as in figure~\ref{fig:Dc_circ}.
It is observed that growths occur more efficiently as $ \psi_3 $ decreases, i.e., as the final direction of the wavevector gets closer to the $ k^2 $-axis.
Indeed, there appears a threshold value $ \psi_3 \sim 10^{-2} $ for growth as implied by eq.~\eqref{eq:psi_th}.
Note also that these figures are to interpolate the top panels of figure~\ref{fig:E_axes}.

\begin{figure}[htbp]
\begin{center}
\includegraphics[scale=0.7]{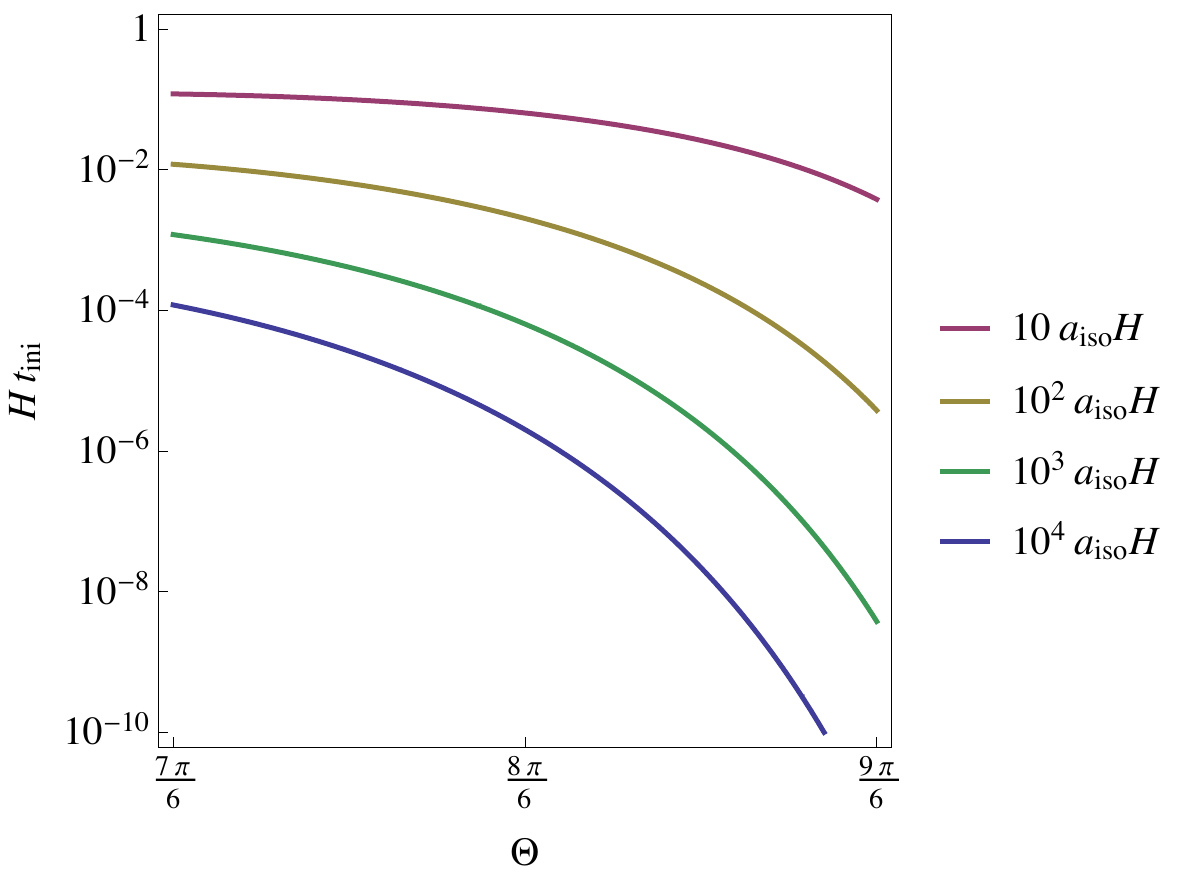}
\end{center}
\caption{\label{fig:C1dist}Boundaries in the $ (\Theta,H t_\mathrm{ini}) $-plane satisfying $ t_\mathrm{ini} = t_\mathrm{stall} $ evaluated for the $ k^2 $-axis-aligned modes with several wavenumbers $ k_2 $\,.
If a combination of the parameters $ (\Theta,H t_\mathrm{ini}) $ lies below a boundary for a given wavenumber $ k_2 $\,, then the corresponding $ k^2 $-axis-aligned mode is supposed to grow and the intensity on $ C_1 $ would look rather uniform.}
\end{figure}

It should be noted here that the situation on $ C_1 $ can change dramatically according to the wavenumber and the background parameters.
As implied by \eqref{eq:t_stall_rapid}, the time scale $ t_\mathrm{stall}^{(\mathrm{fast})} $ near the $ k^2 $-axis on $ C_1 $ ($ \tan\psi_1 \gtrsim 1 $) is a decreasing function of both $ k_0 $ and $ \Theta $\,, the reason for the latter dependence being that the exponent $ q_1^{(\mathrm{fast})} = q_2 $ is an increasing function of $ \Theta $\,.
If $ t_\mathrm{stall}^{(\mathrm{fast})} $ declines to a value as small as $ t_\mathrm{ini} $\,, then amplification of gravitational waves is prohibited near the $ k^2 $-axis.
Therefore, a simple criterion for the growth on $ C_1 $ to occur uniformly may be given by
\begin{equation}
t_\mathrm{ini}
\leq
  t_\mathrm{stall}^{(\mathrm{fast})}(\psi_1 = \pi/2) 
= t_\mathrm{stall}^{(i=2)}\,.
\end{equation}
In figure~\ref{fig:C1dist}, we show the boundaries in the $ (\Theta,H t_\mathrm{ini}) $-plane saturating the above inequality for several wavenumbers, which can be used to discriminate whether the intensity on $ C_1 $ is uniform or not at a corresponding wavenumber.

\subsection{\label{sec:general}Evolution of modes pointing in general directions}

Finally we discuss time evolutions of modes with a wavevector pointing in a general direction.
As indicated in figure~\ref{fig:k_dir}, a wavevector $ \vec k $ rotates from the $ k^1 $- to the $ k^3 $-axis (except on the circumference $ C_3 $).
Since neither of the tensor components of the shear vanishes, the two polarisation modes no longer evolve independently but affect each other.
Hence we shall numerically solve the coupled equations of motion \eqref{eq:eom}.

\begin{figure}[htbp]
\begin{center}
\includegraphics[scale=0.5]{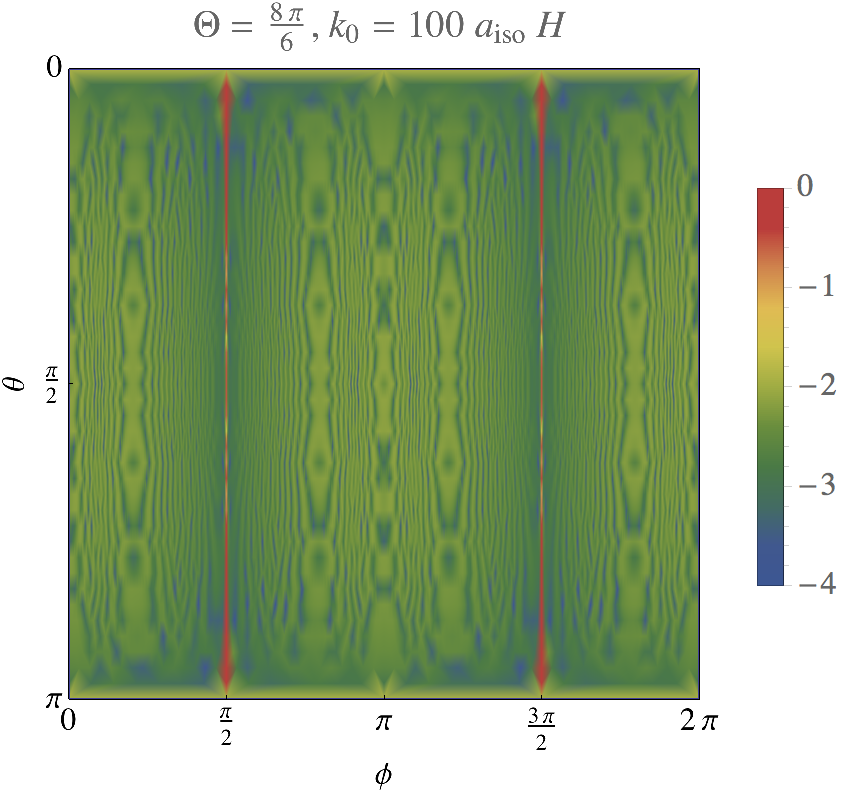}
\includegraphics[scale=0.5]{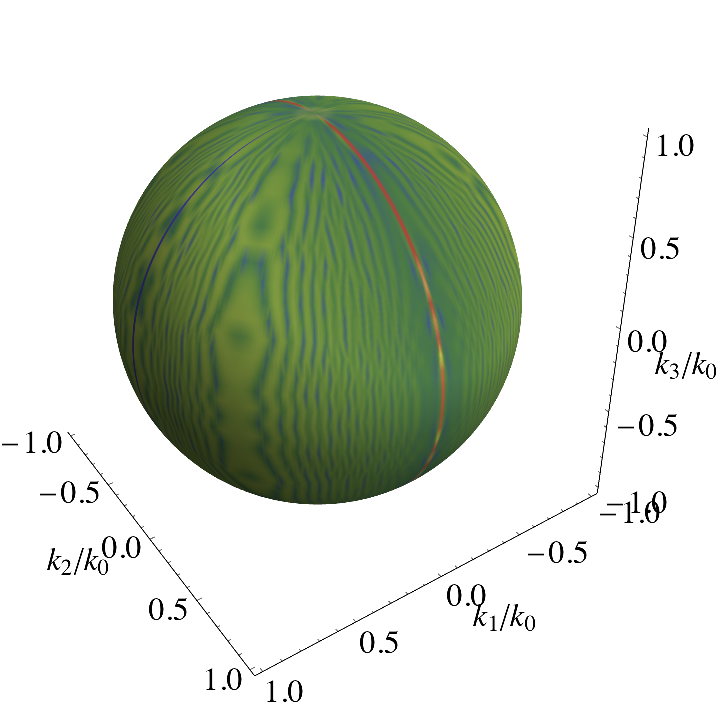} \\
\includegraphics[scale=0.5]{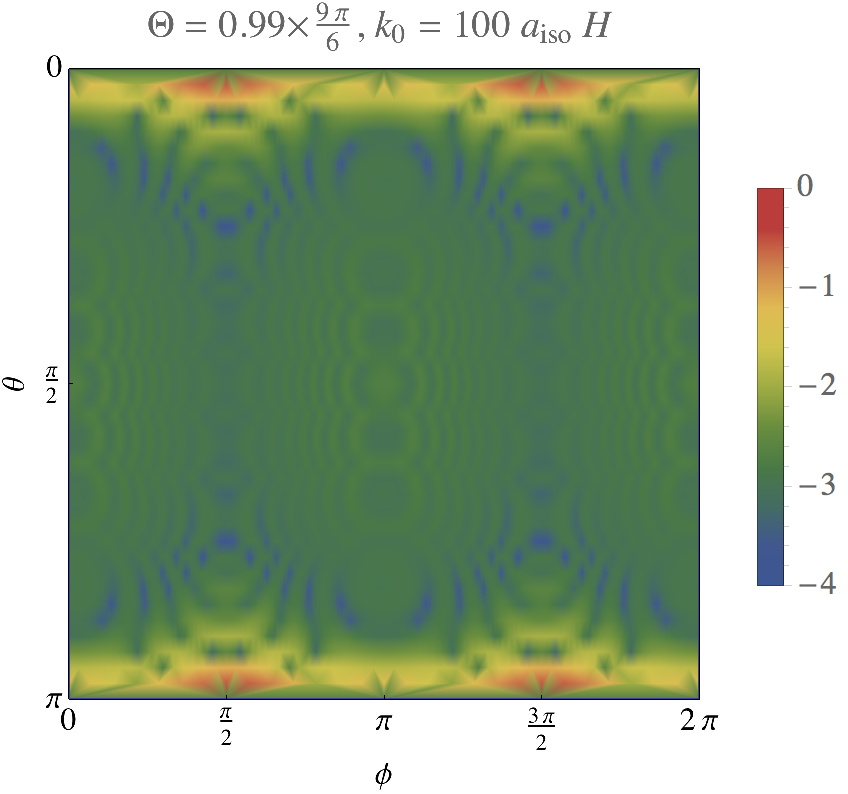}
\includegraphics[scale=0.5]{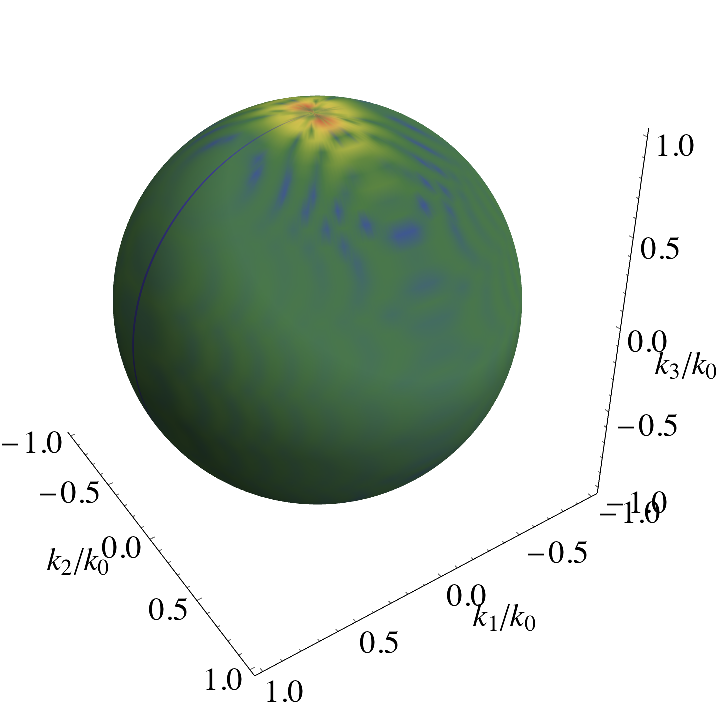}
\end{center}
\caption{\label{fig:I}Qualitative difference in the intensity ratio $ \log_{10}[I(\infty)/I(t_\mathrm{ini})] $ between the cases of $ \Theta = \frac{8\pi}{6} $ (top) and $ 0.99 \times \frac{9\pi}{6} $ (bottom).}
\end{figure}

In figure~\ref{fig:I}, we show all-sky maps of the growth factor for intensity $ I \equiv \sqrt{E_+^2+E_\times^2} $\,, which is invariant under rotations of polarisation basis, for the anisotropy parameters $ \Theta = \frac{8\pi}{6} $ (top) and $ 0.99 \times \frac{9\pi}{6} $ (bottom).
The former represents a highly triaxial configuration while the latter is nearly axisymmetric around the $ k^3 $-axis, resembling the situation investigated in \cite{Gumrukcuoglu:2008gi}.

An obvious feature seen in the top panels is significantly higher intensities along the $ k_1 = 0 $ great circle.
This is a manifestation of the result of the analysis in the previous sections that the growth of gravitational waves only occurs in a narrow range near the slower axis on the circumferences $ C_2 $ and $ C_3 $\,.
Indeed, the width of the belt-shaped region of higher intensities is well estimated by $ \psi_{i=2,3} $ given by eq.~\eqref{eq:psi_th}.
On the other hand, in the bottom panels ($ \Theta = 0.99 \times \frac{9\pi}{6} $) the growth is localised to near the $ k^3 $-axis.

As we shall explain now, this topological property is crucial in determining not only the directions of the principal axes of the early anisotropic expansion but also the degree of anisotropy $ \Theta $ and the initial time of the anisotropic pre-inflationary stage $ t_\mathrm{ini} $ with the aid of figure~\ref{fig:C1dist}.

Let us imagine all-sky observations of gravitational waves at multiple wavelengths become put into practice in the future.
If the gravitational waves from the anisotropic regime are detected at some (long) wavelength and the intensity map is determined to have a topology like the top panels of figure~\ref{fig:I}, then the principal axis of the fastest expansion (in our notation, the $ x^1 $-axis) is determined as the normal direction to the plane containing the great circle of higher intensities.
As for the pre-inflationary parameters, the combination of $ (\Theta,H t_\mathrm{ini}) $ in figure~\ref{fig:C1dist} is restricted to lie below the boundary corresponding to the observation wavelength.

Our analyses further predict that, as one goes from longer wavelengths to shorter, the intensity maps will exhibit a transition from a circle-like one (as depicted in the top panels of figure~\ref{fig:I}) to a localised one (bottom panels) at some critical wavelength.
From the map after the transition, the axis of the slowest expansion (the $ x^3 $-axis) is determined to be the direction pointing the highest intensity region on the celestial sphere, after which the remaining axis (the $ x^2 $-axis) is also determined to be the orthogonal direction to the other two axes.

If such a transition is observed, the values of the parameters $ (\Theta,H t_\mathrm{ini}) $ in figure~\ref{fig:C1dist} should lie on a curve corresponding to the critical wavelength.
Then, once either of the two parameters is given observationally or theoretically, the remainder can be determined.

\section{\label{sec:limit}Constraints from CMB $ B $-mode polarisation}

Our final project is to constrain the gravitational wave background from the anisotropic pre-inflationary phase using the CMB observations.
In the inflationary scenario considered in the present analysis, the total power of primordial tensor perturbations after inflation, denoted as $ \mathcal P_\mathrm T $\,, may be regarded as a sum of the pre-inflationary contribution $ \mathcal P_\mathrm T^\mathrm{pre} $ and the ordinary inflationary contribution $ \mathcal P_\mathrm T^\mathrm{inf} $\,.
Their crucial difference is that, while the inflationary part is expected to be a function of the modulus $ k_0 \equiv \sqrt{k_1^2+k_2^2+k_3^2} $ only, the pre-inflationary part should depend on the components $ (k_1,k_2,k_3) $ differently \cite{Pitrou:2008gk}.
Another complexity arises in general due to different evolutions of the two polarisation modes and their interactions during the anisotropic phase \cite{Pitrou:2008gk}.

In this study, we do not perform a full analysis of the directional variations of the power spectra but rather focus on the axis-aligned modes.
By doing so, we can take advantage of the fact that the two polarisations are decoupled, and hence, the power spectrum after inflation can be represented as a sum of those of the $ + $-mode $ \mathcal P_{\mathrm T+}^\mathrm{pre} $ and the $ \times $-mode $ \mathcal P_{\mathrm T\times}^\mathrm{pre} $ \cite{Gumrukcuoglu:2008gi,Pitrou:2008gk}.
They are not identical in general because the two polarisations evolve differently on an anisotropic background and even their initial values are not necessarily the same.
We here assume that classical gravitational waves already exist at the initial time $ t_\mathrm{ini} $\,.
Denoting the values at $ t_\mathrm{ini} $ as $ \mathcal P_{\mathrm T\lambda}^{\mathrm{pre},\mathrm{ini}} $\,, the power spectra after inflation can be written in terms of the growth factor $ D_\lambda $ as
\begin{equation}
\mathcal P_{\mathrm T\lambda}^\mathrm{pre}(k_i)
= D_\lambda^2\,
  \mathcal P_{\mathrm T\lambda}^{\mathrm{pre},\mathrm{ini}}(k_i)\,,
\end{equation}
where $ k_i $ is the only non-zero component.
If $ \mathcal P_{\mathrm T+}^{\mathrm{pre},\mathrm{ini}}(k_i) $ and $ \mathcal P_{\mathrm T\times}^{\mathrm{pre},\mathrm{ini}}(k_i) $ are of the same order, since $ D_\times \gtrsim D_+ $ in general, we may ignore the $ + $-mode in comparison with the $ \times $-mode at later times.
Then, we can regard the total primordial power spectrum of the pre-inflationary gravitational waves as 
\begin{equation}
\mathcal P_\mathrm T^\mathrm{pre}(k_i)
\approx
  \mathcal P_{\mathrm T\times}^\mathrm{pre}(k_i)
= D_\times^2\,
  \mathcal P_{\mathrm T\times}^{\mathrm{pre},\mathrm{ini}}(k_i)\,,
\label{eq:Ppre}
\end{equation}
whose explicit $ k_i $ dependence, apart from the unknown initial part, is
\begin{equation}
\mathcal P_\mathrm T^\mathrm{pre}(k_i)
\propto
  \left(\frac{k_i}{a_\mathrm{iso} H}\right)^{-2-\frac{2 + 6\Delta_i}{3 (2/3 - q_i)}}
  \left(H t_\mathrm{ini}\right)^{-2\Delta_i}\,
  \mathcal P_{\mathrm T\times}^{\mathrm{pre},\mathrm{ini}}(k_i)\,.
\label{eq:Pprek}
\end{equation}

The upper limit on the primordial tensor power spectrum is given in terms of the tensor-to-scalar ratio $ \mathcal P_\mathrm T/\mathcal P_\mathrm S \equiv r \lesssim 0.27 $ at around $ k_0 = 0.05\,\mathrm{Mpc}^{-1} $\,, where the scalar power spectrum is $ \mathcal P_\mathrm S(k_\mathrm{pivot}) \sim 10^{-9} $ \cite{Ade:2015xua}.
Since it is expected that the $ k^3 $-axis-aligned mode grows the most at shorter wavelengths, we require the primordial tensor power spectrum of the mode travelling along the $ x^3 $-axis to be lower than $ r\,\mathcal P_\mathrm S $ as a conservative limit.
An additional assumption to be made here is that the modes exiting the horizon at $ t = t_\mathrm{iso} $ re-enter the cosmological horizon at sufficiently late times, namely $ a_\mathrm{iso}\,H \sim a(t_0)\,H(t_0) $\,.

\begin{figure}[htbp]
\begin{center}
\includegraphics[scale=0.7]{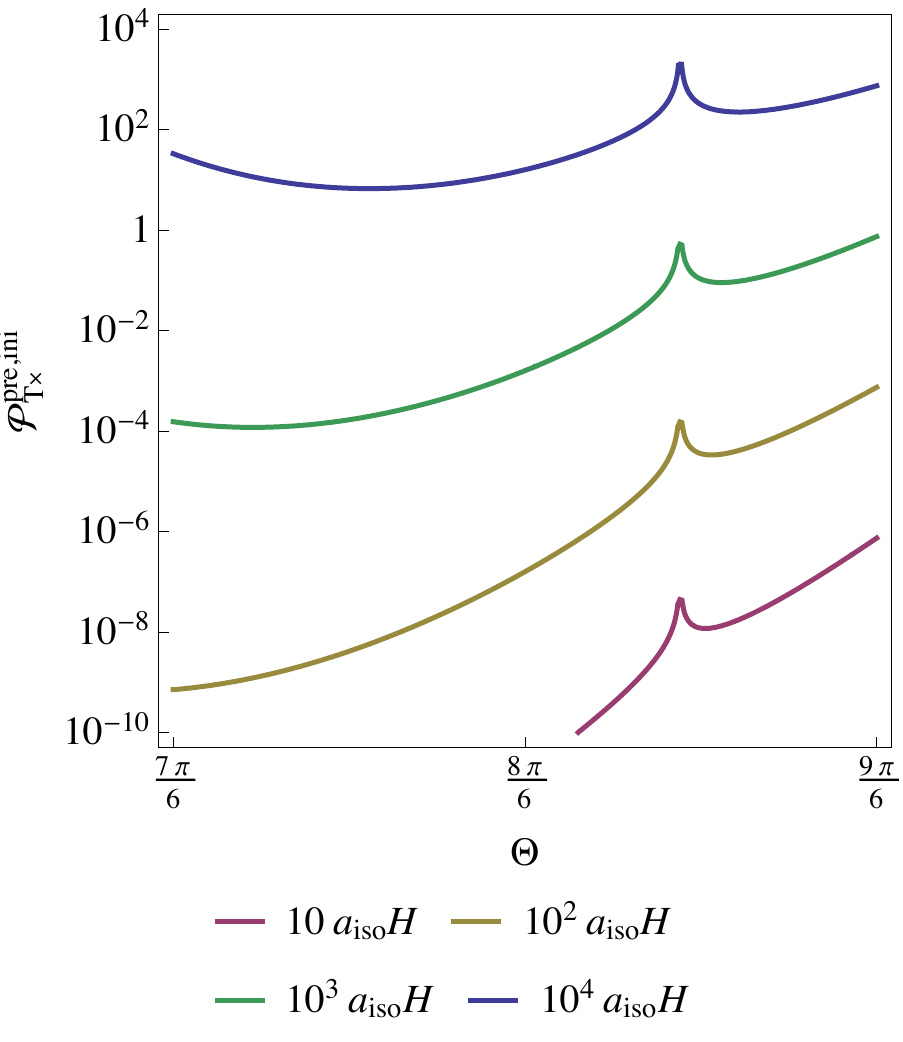}
\includegraphics[scale=0.7]{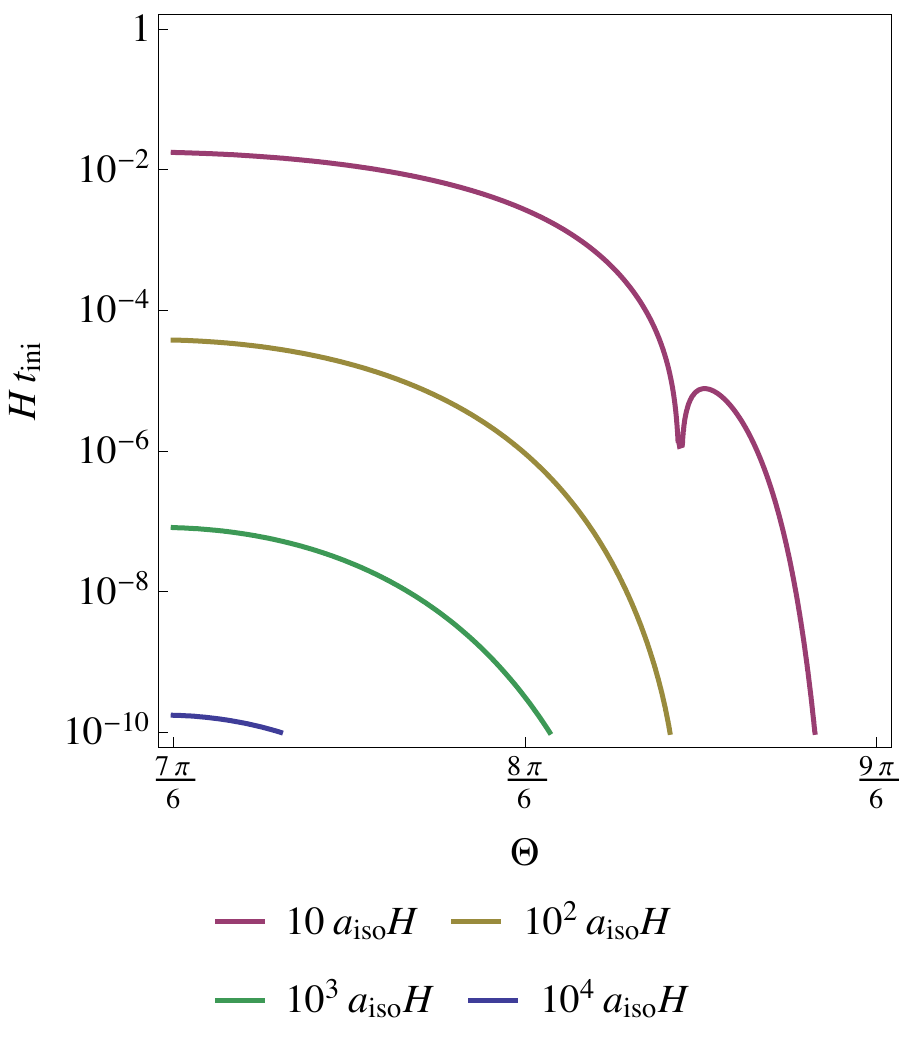}
\end{center}
\caption{\label{fig:const}Left: Upper limits on the initial spectrum of the $ \times $-mode $ \mathcal P_{\mathrm T,\times}^{\mathrm{pre},\mathrm{ini}} $ for wavenumbers $ k_3 = 10\,a_\mathrm{iso}\,H $ (red), $ 10^2\,a_\mathrm{iso}\,H $ (yellow), $ 10^3\,a_\mathrm{iso}\,H $ (green), and $ 10^4\,a_\mathrm{iso}\,H $ (blue).
The initial time is fixed as $ t_\mathrm i = 10^{-5}\,H^{-1} $\,.
Right: Lower limits on the initial time $ H\,t_\mathrm i $ for the same set of wavenumbers.
The initial amplitude of the $ \times $-mode spectrum is fixed as $ \mathcal P_{\mathrm T,\times}^{\mathrm{pre},\mathrm{ini}} = 10^{-8} $\,.}
\end{figure}

In the left panel of figure~\ref{fig:const}, we show upper limits on the $ \times $-mode initial power spectrum of the pre-inflationary gravitational waves $ \mathcal P^{\mathrm{pre},\mathrm{ini}}_{\mathrm T,\times} $ plotted against the anisotropy parameter $ \Theta $\,.
The initial time is fixed to be $ t_\mathrm{ini} = 10^{-5}\,H^{-1} $\,.
The constraint becomes gradually stronger as $ \Theta $ approaches to $ \frac{7\pi}{6} $\,, reflecting the tendency observed in the left panel of figure~\ref{fig:D}.
In the right panel, we show lower limits on the initial time $ t_\mathrm{ini} $\,.
In this figure, we set $ \mathcal P_{\mathrm T\times}^{\mathrm{pre},\mathrm{ini}} = 10^{-8} $\,.

\section{\label{sec:concl}Conclusions}

In this paper, we performed a detailed analysis on the directional variations of a gravitational wave background in a (pre-)inflation model described by the general triaxial Kasner-de Sitter metric, in which the degree of anisotropy is parameterised by an angle parameter $ \Theta $\,.
The purpose of the study was to give some insights into the connection between such gravitational waves and the primordial anisotropies.

We divided the whole analysis into the following three steps.
First, in section~\ref{sec:axes}, we investigated the evolution of gravitational waves whose wavevector is aligned with either of the principal axes of anisotropic expansion.
We clarified that, with our choice of the polarisation basis, the $ \times $-polarisation mode grows substantially while the shear term is giving a dominant contribution to the squared frequency $ \omega_\times^2 $ as given by eq.~\eqref{eq:omega}.
The amount of growth reflects two factors: (i) the slower the expansion along the axis is, the longer $ \omega_\times^2 $ remains negative; (ii) the larger the projected shear component $ \sigma^{(\mathrm T)}_+ $ is, the faster the mode grows.
There was a competition between the $ k^3 $- and $ k^2 $-axis-aligned modes since the former grows for longest while the latter ``sees'' the largest value of $ \sigma^{(\mathrm T)}_+ $\,.
In an analytical manner, we have obtained the explicit parameter dependencies of the growth factors as in eq.~\eqref{eq:Dpi}, clarifying that the $ k^3 $-axis-aligned $ \times $-mode should dominate over the other modes at sufficiently large wavenumbers except for an axisymmetric background with $ \Theta \approx \frac{7\pi}{6} $\,.

Second, in section~\ref{sec:circs}, we extended the analysis to the modes whose wavevector is aligned between two of the principal axes.
The situations are different on each circumference $ C_i $ introduced as in figure~\ref{fig:circ}.
We revealed that the distribution of gravitational-wave intensities on $ C_1 $ can change dramatically according to the background parameters $ \Theta $ and $ t_\mathrm{ini} $ as explained by figure~\ref{fig:C1dist}\,:
when $ \Theta $ is sufficiently close to $ \frac{9\pi}{6} $\,, a value corresponding to one of the axisymmetric limits, and $ H\,t_\mathrm{ini} $ is sufficiently large, the growth of gravitational waves is localised to near the $ k^3 $-axis.
In the opposite case, the growth takes place rather uniformly on $ C_1 $\,.
On the other circumferences $ C_2 $ and $ C_3 $\,, we showed that the growth is localised to near the $ k^3 $- and $ k^2 $-axis, respectively, for wide ranges of parameters.

Third, in section~\ref{sec:general}, we discussed time evolutions of modes with a wavevector pointing in a general direction.
Specifically, we demonstrated all-sky maps of the growth factor of the pre-inflationary gravitational waves for the two anisotropy parameters $ \Theta = \frac{8\pi}{6} $ (triaxial) and $ 0.99 \times \frac{9\pi}{6} $ (nearly axisymmetric) as in figure~\ref{fig:I}.
For $ \Theta = \frac{8\pi}{6} $\,, the growth of gravitational waves occurs in a narrow range near the $ k_1 = 0 $ great circle on the sphere whose width can be well estimated by eq.~\eqref{eq:psi_th}, while for $ \Theta = 0.99 \times \frac{9\pi}{6} $\,, the growth is localised to the $ k^3 $-axis.
Using these results, we argued that the topological properties of the pre-inflationary gravitational waves in future all-sky, multiwavelength observations will provide us a crucial probe for determination of the configuration of the primordial anisotropy, its degree $ \Theta $\,, and the initial time $ t_\mathrm{ini} $\,.

Finally, in section~\ref{sec:limit}, we gave some tentative constraints on the initial amplitude of the pre-inflationary gravitational waves, the anisotropy parameter $ \Theta $\,, and the initial time $ t_\mathrm{ini} $ from the $ B $-mode polarisation of CMB observed by the Planck satellite.

One of possible directions of extending the present work might be inclusion of a scalar degree of freedom which drives isotropisation and inflation instead of a cosmological constant.
Also, it will be meaningful if this sort of analysis can be extended to other Bianchi-type cosmological models.

Perturbation analyses in anisotropic cosmologies will be of particular importance in constraining some kinds of modified theories of gravity.
Among others, the Einstein-Weyl gravity, whose Lagrangian is $ \mathcal L = \frac{1}{2}\,R - \frac{\gamma}{4}\,C_{\alpha\beta\gamma\sigma}\,C^{\alpha\beta\gamma\sigma} $\,, has a special property that it admits general Einstein spaces, including KdS, as exact solutions.
We will come back to this issue in future publications.

\acknowledgments

Y.F.\ and Y.S.\ are grateful to Nathalie Deruelle for helpful comments and stimulating discussions.
They also acknowledge a financial support and hospitality received from Yukawa Institute for Theoretical Physics, Kyoto University, at an early stage of this study.
This work was in part supported by JSPS KAKENHI Grants No.~26800115 and No.~JP16K17675 (Y.S.).

\bibliographystyle{JHEP}
\bibliography{aniso}

\end{document}